\documentclass[a4paper,11pt]{article}
\usepackage{jcappub}
\usepackage{hyperref}
\usepackage{cleveref}
\usepackage{subcaption}
\usepackage[varg]{txfonts}
\usepackage{graphicx} 
\usepackage{gensymb}
\usepackage{amsmath}
\usepackage{multirow}

\title{Is cosmic birefringence due to dark energy or dark matter? Simulation-based inference}

\emailAdd{fcarralo@sissa.it}

\author[a,b]{Florie Carralot,}
\author[c]{Patricia Diego-Palazuelos,}
\author[c]{Adriaan\ J.\ Duivenvoorden,}
\author[c,d,e]{Eiichiro Komatsu,}
\author[a,f,g]{Nicoletta Krachmalnicoff,}
\author[a,f,g]{Carlo Baccigalupi}

\affiliation[a]{International School for Advanced Studies (SISSA), Via Bonomea 265, 34136, Trieste, Italy}
\affiliation[b]{Universita di Trento, Dipartimento di Fisica, Via Sommarive 14, 38123, Trento, Italy}
\affiliation[c]{Max Planck Institute for Astrophysics, Karl-Schwarzschild-Str. 1, D-85748 Garching, Germany}
\affiliation[d]{
Ludwig-Maximilians-Universit\"at M\"unchen, Schellingstr.\ 4, 80799 M\"unchen, Germany
}
\affiliation[e]{
Kavli Institute for the Physics and Mathematics of the Universe (Kavli IPMU, WPI), UTIAS, The University of Tokyo, Chiba, 277-8583, Japan
}
\affiliation[f]{INFN Sezione di Trieste, via Valerio 2, 34127 Trieste, Italy}
\affiliation[g]{IFPU, Via Beirut, 2, 34151 Grignano, Trieste, Italy}
\abstract{Simulation-based inference (SBI) is a powerful inference technique for cases where the exact functional form of the likelihood is not known. A prime example is the likelihood of cross-correlation power spectra of the cosmic microwave background (CMB) fields at low multipoles, $\ell\lesssim 10$. In this paper, we investigate a parity-violating cross-correlation between $E$- and $B$- mode polarization fields using SBI. The $EB$ correlation at low $\ell$ is essential to distinguish between possible axion dark energy and dark matter interpretations of `cosmic birefringence', a rotation of the plane of linear polarization of the CMB, recently reported from \textit{WMAP}, \textit{Planck}, and Atacama Cosmology Telescope data. We use neural likelihood estimation to infer the likelihood of the $EB$ correlation at low $\ell$ and show that it is highly non-Gaussian. We then employ neural posterior estimation to constrain the scalar field mass ($m_\phi$), the cosmic birefringence amplitude ($g\phi_\mathrm{in}/2$), and the instrumental miscalibration angle ($\alpha$), from simulated datasets. We find that the posterior on $m_{\phi}$ shows two regimes, with a transition marked by $10^{-32}$ eV, highlighting a strong sensitivity to the scale dependence of cosmic birefringence. To quantify this behavior, we compute the probability $p(m_{\phi} < 10^{-32}$\,eV) for various fiducial values of $m_{\phi}$. We find that $\alpha$ and the contribution of lensed $B$ modes ultimately limit our ability to exclude the dark energy scenario fully. }


\begin{document}
\maketitle

\flushbottom

\section{Introduction}

The standard model of cosmology, known as the $\Lambda$ Cold Dark Matter ($\Lambda$CDM) model, is grounded in fundamental physical laws that are symmetric under parity transformations. In particular, it does not predict $TB$ and $EB$ cross-correlations between the parity-even ($T$, $E$) and parity-odd ($B$) anisotropies of the cosmic microwave background (CMB), where $T$ denotes temperature, and $E$ and $B$ represent the parity-even and parity-odd components of the linear polarization field~\cite{Kamionkowski:1996ks,Seljak:1996gy}. Detecting such correlations from CMB observations would provide strong evidence for parity-violating mechanisms in the Universe~\cite{1999PhRvL..83.1506L}, offering a window into new physics~\cite{newphysCMB,Nakai:2023zdr}.

Parity-violation in the CMB could arise from a pseudo-scalar field, $\phi$, potentially associated with axionlike particles~\cite{Marsh_2016}, that changes sign under parity. This field can couple with the electromagnetic tensor, $F_{\mu\nu}$, and its dual, $\tilde{F}^{\mu\nu}$, through a \textit{Chern Simons} term \cite{PhysRevLett.38.301,Cst1,Sikivie:1983ip}, introducing an additional element in the Lagrangian density: $\mathcal{L} \subset -\frac{1}{4}\phi gF_{\mu\nu}\tilde{F}^{\mu\nu}$. When $\phi$ is dynamical, this term induces different phase velocities for left- and right-handed photons, resulting in a net rotation of the polarization plane of the CMB by an angle $\beta$ (e.g., see review~\cite{newphysCMB}). This effect, referred to as \textit{cosmic birefringence}, gives rise to nonzero $EB$ and $TB$ parity-violating correlations. In the rest of the paper, we focus exclusively on the $EB$ correlation as it has a higher signal-to-noise ratio than $TB$ \cite{Sullivan:2025btc}. 

Recent constraints on cosmic birefringence from the analysis of \textit{Planck} data~\cite{Minami:2020odp, 2022PhRvL.128i1302D, Eskilt:2022wav, Sullivan:2025btc, Remazeilles:2025wzd}, its joint analysis with \textit{WMAP}~\cite{2022PhRvD.106f3503E, Cosmoglobe:2023pgf}, and the Atacama Cosmology Telescope (ACT) data release 6~\cite{AtacamaCosmologyTelescope:2025blo, 2025arXiv250913654D}, have reported consistent positive $EB$ signals described by $\beta \approx 0.3 \degree$. Independently, the statistical significance of these results ranges from $2.4$ to $3.6\,\sigma$, reaching a $4.6\,\sigma$ confidence level when combined~\cite{2025arXiv250913654D}. The main impediment these measurements face for observing the $EB$ power spectrum is the systematic rotation produced by miscalibration of instrumental polarization angles, $\alpha$. The uncertainty in the orientation of the detectors' polarization plane and nonidealities in the optics used for the experiments induce spurious nonzero $EB$ and $TB$ correlations, which can be confused with the cosmic birefringence signal~\cite{QUaD:2008ado, 2009PhRvD..79j3002M}. A miscalibration of $\alpha$ can be corrected for through the use of precise models of the telescope optics~\cite{Murphy:2024fna, 2025arXiv250913654D} or polarized Galactic emission~\cite{Minami:2019ruj, 2022PhRvL.128i1302D}. Although confirmation from experiments using artificial polarization sources as calibrators instead of relying on models of optics or Galactic polarization, such as BICEP3~\cite{BICEPKeck:2024cmk}, CLASS~\cite{Coppi:2025fmt}, and the Simons Observatory (SO)~\cite{2023RScI...94l4502M}, is still missing, it is compelling that independent data sets and calibration methodologies have yielded consistent values for $\beta$. 

The nature of the field responsible for the observed signal remains, however,  uncertain: does $\phi$ act as dark energy, or could it instead correspond to a fraction of dark matter? If the mass is $m_{\phi} \lesssim H_0 \approx 10^{-33}$\,eV, the field is frozen across cosmological timescales and behaves as dark energy today. This evolution results in the $EB$ cross-correlation power spectrum that is proportional to the difference between the $EE$ and $BB$ auto power spectra at all angular scales. In contrast, when $10^{-33}\,\mathrm{eV}<m_{\phi} < 10^{-28}$\,eV, $\phi$ evolves during the reionization epoch, leading to a different angular dependence of the $EB$ power spectrum \cite{Sherwin:2021vgb,Nakatsuka_2022}. This is characteristic of a field that constitutes a fraction of dark matter.  Hence, the full shape of the $EB$ power spectrum offers crucial insights into the time evolution of the field, allowing constraints to be placed on its mass, and helps to disentangle the cosmological signal from instrumental miscalibration of $\alpha$, as shown in \cite{Sherwin:2021vgb,Nakatsuka_2022}. Specifically, detecting the $EB$ power spectrum that is incompatible with the difference between the $EE$ and $BB$ power spectra would effectively rule out the dark energy scenario. We emphasize that observing such properties from the $EB$ power spectrum requires access to the largest angular scales ($\ell < 20$). \cite{Ballardini:2025apf} found no evidence for a scale-dependent birefringence signal on the $\ell>42$ angular scales of \textit{Planck} $EB$. The only upcoming experiment capable of measuring both reionization ($\ell\lesssim10$) and recombination ($50\lesssim\ell\lesssim200$) bumps of the CMB power spectrum is the \textit{LiteBIRD} satellite~\cite{2023PTEP.2023d2F01L}, whose launch is planned for the 2030s. Its complementarity with ground-based CMB experiments that are sensitive to smaller angular scales, such as SO~\cite{2019JCAP...02..056A}, ACT~\cite{2020JCAP...12..045C}, and the South Pole Telescope~\cite{SPT-3G:2021eoc}, could provide a complete picture of cosmic birefringence and $\phi$'s dynamics. 

To fully capture the evolution of the $\phi$ field and infer its underlying parameters from the $EB$ power spectrum, it is essential to extract information from both reionization and recombination bumps. A major challenge in using the $EB$ power spectrum for parameter inference is that its statistical distribution deviates significantly from Gaussianity at low multipoles, making standard likelihood approaches unsuitable~\cite{2006MNRAS.372.1104P}. This complication arises partly from our choice of using only the $EB$ power spectrum, rather than exploiting the full set of polarization spectra ($EE,\,BB,\,EB$), for which the form of the likelihood is known \cite{Gerbino:2019okg}. The motivation for focusing only on $EB$ is that $EE$ and $BB$ are typically more susceptible to foreground contamination, instrumental systematic effects, and correlated noise. However, this choice comes with an intractable likelihood function whose form can become even more complex in the presence of Galactic emission and partial sky coverage. Consequently, it is essential to use alternative approaches capable of performing parameter inference without direct access to the likelihood.

In this paper, we employ \textit{simulation-based inference} (SBI) \cite{2025arXiv250812939D,2020PNAS..11730055C} to constrain the scalar field mass $m_{\phi}$, the cosmic birefringence amplitude $g\phi_{\mathrm{in}}/2$, and the polarization angle miscalibration $\alpha$ using the $EB$ power spectrum. 
SBI refers to a class of methods designed to infer posterior distributions over model parameters using synthetic data generated from a forward model, often referred to as \textit{simulator}. A key advantage of SBI is that it does not require an explicit form of the likelihood distribution. While several subcategories of SBI methods exist \cite{ABC}, this analysis focuses on a neural network-based approach that learns probability distributions relating data and parameters. 

This paper is organized as follows. \Cref{axion_dynamics} presents the characteristics of the $EB$ power spectrum under different field dynamics. The SBI framework is introduced in \cref{SBI}, and the implementation of the simulator used to build the training dataset is presented in \cref{sims}. \Cref{Resultats} presents the constraints on $m_{\phi}$, $g\phi_{\mathrm{in}}/2$, and $\alpha$ for different SBI approaches. A discussion relating the main results and possible improvements to this study is given in \cref{Discussion}.

\section{$EB$ power spectrum from scalar field dynamics}\label{axion_dynamics}

The equation of motion of a homogeneous scalar field $\phi$, evolving in a quadratic potential $V(\phi) = {m_{\phi}^{2}{\phi}^{2}/2}$, is given by the Klein-Gordon  equation \cite{Kolb:1990vq}:
\begin{equation}\label{KG}
    \ddot{\phi} + 3H\dot{\phi} + {m_{\phi}}^{2}\phi = 0 \, ,
 \end{equation}
where $\dot{x}$ denotes derivatives with respect to the physical time. When $m_{\phi} \ll H(t)$, the friction term $3H\dot{\phi}$ dominates and the field evolution is overdamped by cosmic expansion. The field is therefore frozen and evolves slowly compared to the expansion rate. In this regime, the scalar field corresponds to a fluid with equation of state $w=-1$, and thus behaves as a dark energy component. In contrast, when $m_{\phi} \gg H(t)$, the mass term dominates in \cref{KG}, and the field oscillates around the minimum value of its potential. In this case, the field behaves as a pressureless fluid (time-averaged $\langle w\rangle\approx0$) contributing to a fraction of the total dark matter density~\cite{Rogers:2023ezo}. In particular, axionlike particles would act as dark energy today if $m_{\phi}\lesssim H_{0} \approx {10}^{-33}$\,eV, while constituting a fraction of the dark matter today if $m_{\phi} >  {10}^{-33}$\,eV. 

The mass governs the field evolution, leading to distinct angular-scale signatures in the $EB$ power spectrum \cite{Sherwin:2021vgb,Nakatsuka_2022}. Specifically, the cosmic birefringence angle at a certain conformal time, $\eta$, is expressed as a difference in the field values between the emission and observation \cite{Carroll:1991zs,Harari:1992ea}:  
\begin{equation}\label{biref_angle}
    \beta(\eta) = \frac{g\phi_{\mathrm{in}}}{2}[f(\eta_0) - f(\eta)] \, ,
\end{equation}
where $g$ represents the coupling constant of $\phi$ and photons, $f(\eta) = \phi(\eta)/\phi_{\mathrm{in}}$, and $\eta_0$ corresponds to the conformal time today. In \cref{biref_angle}, the constant $g \phi_{\mathrm{in}}/2$ can be interpreted as a `cosmic birefringence amplitude' (in degrees, after unit conversion of the dimensionless $g\phi_\mathrm{in}$). When the field acs as dark energy, \cref{biref_angle} gives an isotropic cosmic birefringence angle that uniformly multiplies all angular scales of the $EB$ power spectrum~\cite{Sherwin:2021vgb}.

\begin{figure}
    \centering
    \includegraphics[width=\textwidth]{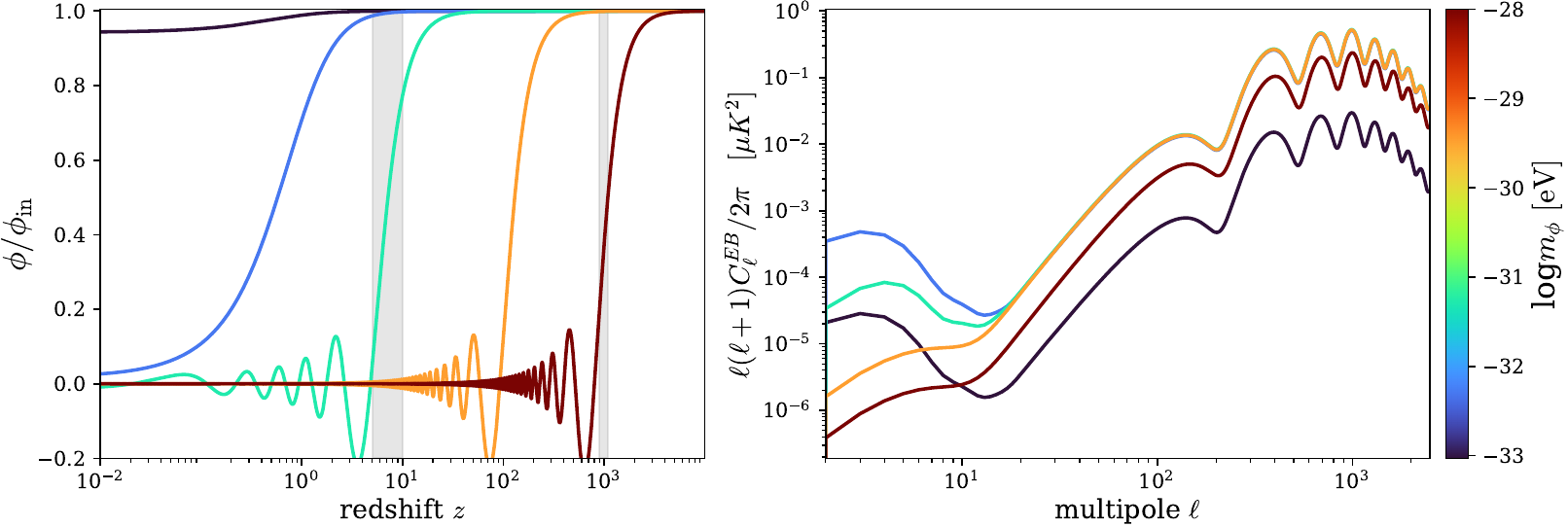}
    \caption{\textit{Left}: Evolution of $\phi$ (normalized at its initial value) as a function of redshift for different masses. The shaded areas represent the reionization ($z \sim 10$) and recombination  ($z \sim 1100$) epochs. \textit{Right}:  The corresponding CMB $EB$ power spectra for the same masses. For all spectra, the cosmic birefringence amplitude is $g \phi_{\mathrm{in}}/2 = 0.35^\circ$.}
    \label{axion_potential}
\end{figure}

The evolution of $\phi$ for different masses and the corresponding $EB$ power spectra are presented in \cref{axion_potential}. For $m_{\phi} \lesssim 10^{-32}$\,eV, the field does not evolve during recombination and reionization epoch and remains fixed at its initial value, so that $\phi_{\mathrm{rec}} = \phi_{\mathrm{reio}} = \phi_{\mathrm{in}}$. In this case, the $EB$ power spectrum is given by $\frac{1}{2}\sin(4\beta)(C_{\ell}^{EE} - C_{\ell}^{BB})$, with $\beta= g\phi_{\mathrm{in}}\Delta f/2$ and $\Delta f = f(\eta_0) - f(\eta)$. Assuming $\beta$ to be small and  $C_{\ell}^{BB} \ll C_{\ell}^{EE}$, it implies $C_{\ell}^{EB} \approx 2\beta\, C_{\ell}^{EE}$. A distinct signature of dark energy fields is therefore finding an $EB$ spectrum with a similar shape as $EE$. In the extreme limit where $m_\phi \ll 10^{-32}$\,eV (dark blue), the field remains effectively frozen up to the present time, leading to  $\beta \propto \Delta f\to 0$ according to \cref{biref_angle}. Consequently, the characteristic $EE$-like shape of the $EB$ power spectrum is preserved, but the signal strength diminishes for smaller masses.

For larger masses, $m_{\phi} \gtrsim 10^{-32}$\,eV, the field starts evolving during (or even before) the reionization epoch, and enters an oscillatory phase around the minimum of the potential. In this case, $\beta(\eta)$ is smaller at reionization than at recombination. 
This produces a suppression of the `reionization bump' in the $EB$ signal at $\ell \lesssim 10$. Therefore, the cosmic birefringence effect becomes scale-dependent for fields that contribute to dark matter \cite{Finelli:2008jv,Nakatsuka_2022}. Detecting such a scale-dependent cosmic birefringence effect would indicate a complex time evolution of $\phi$, and, therefore, rule out the dark energy origin of the signal. Note that even when the field has completely vanished during reionization, a small reionization bump remains in the ${EB}$ spectrum: the $E$ modes converted into $B$ modes during recombination remain correlated with the large-scale $E$ modes produced during the reionization epoch \cite{Nakatsuka_2022}. When the mass exceeds $m_{\phi} \gtrsim 10^{-29}$\,eV (brown), the field starts evolving before or during recombination, and its oscillations result in a cancelation of the birefringence effect, leading to an overall suppression of $C_\ell^{EB}$~\cite{Finelli:2008jv,Murai:2024yul}.

As shown in \cite{Nakatsuka_2022}, the threshold mass that marks the distinction between scale-independent and scale-dependent cosmic birefringence effect is $\sim 10^{-32}$\,eV. The goal of this work is therefore to probe the nature of $\phi$ by identifying any potential scale dependence in the cosmic birefringence signal that could rule out the dark energy interpretation of the $EB$ spectrum, rather than placing strong constraints on its mass.

\section{Simulation-based inference}\label{SBI}

\subsection{Inference networks}\label{inference_networks}

In this paper, we consider the SBI \cite{Tejero-Cantero:2020sbi, Boelts:2025sbiReloaded} method that makes use of neural networks to learn probability distributions relating simulated data, $x$, and model parameters, $\theta$. These \textit{inference networks} rely solely on the prior knowledge of $\theta$ and a model to generate simulations, called the \textit{simulator}. By sampling parameters from the prior, $\theta \sim p(\theta)$, and producing synthetic data with the simulator, $x \sim p(x|\theta)$, SBI is capable of learning different probability distributions relating $x$ and $\theta$, depending on the choice of one of these algorithms: \textit{neural posterior estimation} (NPE), which directly estimates the posterior distribution, $p(\theta|x)$~\cite{2019arXiv190507488G,Papamakarios:2016ctj}; \textit{neural likelihood estimation} (NLE), which estimates the likelihood, $p(x|\theta)$~\cite{2018arXiv180507226P,Glockler:2022vnb}; and \textit{neural ratio estimation} (NRE), which estimates the likelihood-to-evidence ratio, 
$p(x|\theta)/p(x)$~\cite{2020arXiv200203712D}. 

The NLE framework can be particularly advantageous as it enables a straightforward combination with other observational constraints through simple likelihood multiplication. However, obtaining the posterior distribution still requires an additional Monte Carlo sampling step. In contrast, the NPE approach directly approximates the  posterior distribution over the model parameters, $p(\boldsymbol{\theta}|\boldsymbol{x}) = p(\theta_1, \theta_2, \ldots, \theta_N|x_1,x_2,...,x_N)$.\footnote{Throughout the paper, bold symbols designate vector quantities.} 

In the case of multiple observational (test) datasets, one may favor an \textit{amortized} inference approach \cite{Radev:2020bayesflow,Gloeckler:2023adversarial} that learns a single inference model for all available observations supported by the prior. This method  amortizes the computational cost since it can reproduce an accurate posterior for any new dataset, without needing to re-train the model. It is particularly advantageous when dealing with multiple observations, but can be suboptimal in terms of simulation cost as it often requires a large number of simulations.

An efficient method allowing a reduction of the simulation budget is \textit{sequential} inference \cite{2018arXiv180507226P, 2020arXiv200203712D}. Sequential neural posterior estimation (SNPE) aims to estimate the posterior distribution over parameters given a specific observation  $\boldsymbol{x}^{\mathrm{obs}}$, through an iterative process. In the first round, parameters $\boldsymbol{\theta}$ are sampled from a proposal distribution $\tilde{p}(\boldsymbol{\theta})$, initially the prior, and simulations are generated. A neural network is trained to approximate the posterior, which is then conditioned in $\boldsymbol{x}^{\mathrm{obs}}$  and serves as a proposal for the next round. In each subsequent round, parameters are sampled from the updated proposal, generating new simulations that focus on regions of the parameter space more consistent with the observation. The neural network is retrained using the new data, and the proposal is replaced with the latest posterior estimate. This sequential refinement allows the posterior to become increasingly accurate by focusing on relevant regions of the parameter space, for a given observation. For this reason, the SNPE method is expected to achieve convergence to the true posterior with fewer simulations with respect to the amortized case. However, because the posterior estimate is tailored to a single observation, the inference is no longer amortized and therefore does not guarantee an accurate posterior estimate for another observation. As a result, the neural network must be retrained for each new observation. 

In this work, we explore amortized and sequential inference in \cref{am_inf,SNPE}, respectively, using the tools provided by the \texttt{SBI}\footnote{\url{https://sbi.readthedocs.io/en/latest/}} software package.

\subsection{Normalizing flows}

The algorithms presented in \cref{inference_networks} define the target probability distribution, but various mechanisms, known as \textit{neural density estimators}, can approximate probability distributions. In this work, we employ normalizing flows (NFs)~\cite{NFlows} that transform a simple base distribution into a more complex target distribution through a sequence of invertible and differentiable transformations. The key idea behind NFs is that, despite the simplicity of the base distribution (e.g., a standard Gaussian) and the potential complexity of the target distribution (e.g., non-Gaussian or multi-modal), one can construct a sequence of bijective transformations that maps the base distribution to the target. 

We call \textbf{x} our data vector, and \textbf{u} a vector sampled from a simple base distribution, $\textbf{u} \sim p_u(\textbf{u})$. Consider a composition $T$ of invertible and differentiable transformations, such that $\textbf{x} = T(\textbf{u})$ and $\textbf{u} = T^{-1}(\textbf{x})$. This construction allows us to express the density of \textbf{x} in terms  of the base distribution $p_u$ and the composition of transformations $T$: 
\begin{equation}\large\label{density_nf}
    p(\boldsymbol{x}) = p_u(\boldsymbol{u})\left|\det\left(\frac{\partial T^{-1}}{\partial \boldsymbol{x}}\right)\right| = p_u(T^{-1}(\boldsymbol{x}))\left|\det\left(\frac{\partial T^{-1}}{\partial \boldsymbol{x}}\right)\right|.
\end{equation}

We assume a \textit{flow-based model}, $q_{\boldsymbol{\lambda}}(\boldsymbol{\theta}|\boldsymbol{x})$, that begins with a simple base distribution and applies a sequence of invertible transformations that progressively map it to the unknown target distribution. The model is fully specified by a set of parameters $\boldsymbol{\lambda} = \{{\boldsymbol{\gamma},\boldsymbol{\rho}}\}$, where $\boldsymbol{\rho}$ and $\boldsymbol{\gamma}$ are associated to the transformations and base distribution features, respectively. During the training phase, the $\boldsymbol{\lambda}$ parameters are adjusted to minimize the discrepancy between the flow-based model and the target. In the context of NPE, this consists of fitting a flow-based model to the posterior distribution, $p(\boldsymbol{\theta}|\boldsymbol{x})$, by minimizing the forward Kullback-Leibler (KL) divergence, which quantifies the difference between  $p(\boldsymbol{\theta}|\boldsymbol{x})$ and $q_{\boldsymbol{\lambda}}(\boldsymbol{\theta}|\boldsymbol{x})$:
\begin{align}\label{KL}
    \mathcal{L}(\boldsymbol{\lambda}) 
    &= \mathbb{E}_{p(\boldsymbol{x})} \ D_{\mathrm{KL}}\left[ p(\boldsymbol{\theta}|\boldsymbol{x}) \,\middle\|\, q_{\boldsymbol{\lambda}}(\boldsymbol{\theta}|\boldsymbol{x}) \right]  \\
     &= \int  p(\boldsymbol{x}) \int  \ p(\boldsymbol{\theta}|\boldsymbol{x}) \left[ 
        \log \frac{p(\boldsymbol{\theta}|\boldsymbol{x})}{q_{\boldsymbol{\lambda}}(\boldsymbol{\theta}|\boldsymbol{x})} 
    \right] \mathrm{d}\boldsymbol{\theta} \mathrm{d}\boldsymbol{x} \nonumber   \\
     &= \int p(\boldsymbol{x},\boldsymbol{\theta})  \left[ 
        \log p(\boldsymbol{\theta}|\boldsymbol{x}) - \log {q_{\boldsymbol{\lambda}}(\boldsymbol{\theta}|\boldsymbol{x})} 
    \right] \mathrm{d}\boldsymbol{\theta} \mathrm{d}\boldsymbol{x} \nonumber   \\
    &= -\mathbb{E}_{p(\boldsymbol{x},\boldsymbol{\theta})} 
        \log q_{\boldsymbol{\lambda}}(\boldsymbol{\theta}|\boldsymbol{x}) + \mathrm{const} \nonumber.
\end{align}
where $p(\boldsymbol{x},\boldsymbol{\theta}) = p(\boldsymbol{\theta}|\boldsymbol{x}) p(\boldsymbol{x})$. The flow parameters are optimized by minimizing the loss function in the final expression of \cref{KL}. The constant term does not affect the optimization process, as it is independent of $\boldsymbol{\lambda}$.
Analogously, the loss function in the case of NLE is $ -\mathbb{E}_{p(\boldsymbol{x},\boldsymbol{\theta})} \log q_{\boldsymbol{\lambda}}(\boldsymbol{x}|\boldsymbol{\theta})$. 
        
By changing variables, both the flow-based model and the posterior distributions are expressed in terms of $\mathbf{u}$ and the transformation $T$. Then, \cref{KL} becomes: 
\begin{equation}\label{NF_params_likelihood}
    \mathcal{L}(\boldsymbol{\lambda}) =  -\mathbb{E}_{p(\boldsymbol{x},\boldsymbol{\theta})} \big[ \log p_{\boldsymbol{u}}(T^{-1}(\boldsymbol{x,\boldsymbol{\lambda}}),\boldsymbol{\gamma}) + \log |\det J_{T^{-1}}(\boldsymbol{x,\boldsymbol{\lambda}})|\big],
\end{equation}
where $J_{T^{-1}}$ denotes the Jacobian of $T^{-1}$. \Cref{NF_params_likelihood} highlights that this estimate does not require evaluations of the target distribution. 

For the normalizing flow, we use a Masked Autoregressive Flow (MAF) \cite{MAF} network architecture. The MAF network is part of class of \textit{autoregressive flow} models that decompose a joint probability density as a product of one dimensional conditional densities $p(\boldsymbol{x}) = \prod_i p(x_i|\boldsymbol{x}_{1:i-1})$.  The autoregressive nature (i.e., each $x_{i}$ depends only on the previous inputs $\boldsymbol{x}_{1:i-1}$) ensures that the Jacobian in \cref{density_nf} becomes triangular, which means that its determinant can be evaluated efficiently as the product of the diagonal terms.

\section{Simulator}\label{sims}

In this section, we outline the simulation procedure used to generate the SBI training dataset to jointly constrain $m_{\phi}$, $g \phi_{\mathrm{in}}/2$ and $\alpha$. 

We start by computing $N_{\mathrm{sim}}$ CMB power spectra, denoted $C_{\ell}^{XY}$ (with $X,Y=E,B$), for different values of $m_{\phi}$ and $g\phi_{\mathrm{in}}/2$ and over a multipole range from $\ell_{\mathrm{min}} = 2$ to $\ell_{\mathrm{max}} = 2500$. These calculations are performed with the \texttt{birefCLASS}~\cite{Nakatsuka_2022,Murai:2022zur,Naokawa:2023upt} code, which solves the Boltzmann equations coupled with the dynamics of $\phi$, assuming that the energy density of $\phi$ is negligible (i.e., $\phi$ is a spectator field). We adopt the \textit{Planck} best-fit cosmological parameters \cite{Planck:2018vyg}, fixing the optical depth at reionization to $\tau = 0.054$ and assuming no primordial tensor $B$ modes (tensor-to-scalar ratio of $r=0$).
Due to its high computational cost, \texttt{birefCLASS} is not suitable for the multi-round inference loops described in \cref{SNPE}. Therefore, we employ the \texttt{CosmoPower}\footnote{\url{https://alessiospuriomancini.github.io/cosmopower/}}\cite{2022MNRAS.511.1771S} emulator to learn the underlying structure of the CMB power spectra under varying cosmological parameters, allowing for highly efficient generation of new spectra. Further details on the emulator are presented in \cref{Emulator_validation}.

A miscalibration angle $\alpha$ is then propagated to the previously computed  $C_{\ell}^{EE}$, $C_{\ell}^{BB}$ and $C_{\ell}^{EB}$ power spectra, according to the following transformation relationships \cite{Pagano:2009kj}:
\begin{equation}
\begin{cases}
{C}_{\ell}^{EE,\mathrm{obs}} = \cos^2(2\alpha)C_{\ell}^{EE} + \sin^2(2\alpha) C_{\ell}^{BB} - \sin(4\alpha)C_{\ell}^{EB}   \\
{C}_{\ell}^{BB,\mathrm{obs}} = \cos^2(2\alpha)C_{\ell}^{BB} + \sin^2(2\alpha)C_{\ell}^{EE} + \sin(4\alpha) C_{\ell}^{EB} \\
{C}_{\ell}^{EB,\mathrm{obs}} = \frac{1}{2}\sin({4\alpha})[C_{\ell}^{EE}-C_{\ell}^{BB}] + \cos(4\alpha){C}_{\ell}^{EB}. \\ 
\end{cases}\label{miscal}
\end{equation}

\Cref{miscal} yields $N_{\mathrm{sim}}$ of $EE, BB$ and $EB$ power spectra for various combinations of $m_{\phi}$, $g\phi_{\mathrm{in}}/2$ and $\alpha$. As required for the application of SBI, these three parameters are sampled from specific prior distributions. We draw $m_{\phi}$ from a uniform distribution as $\log m_{\phi} \sim \mathcal{U}(-34,-28)$, to cover a broad mass range that encompasses both dark energy and dark matter scenarios. The constraints are set on the logarithm of $m_{\phi}$ with base 10 to avoid extremely small numbers and prevent numerical instabilities. We employ a Gaussian prior, $g\phi_{\mathrm{in}}/2 \sim \mathcal{N}(0,0.3\degree)$, for $g\phi_{\mathrm{in}}/2$ based on observational hints of a potential deviation from $\beta=0$ in the range of  $0.2$ to $0.35\degree$~\cite{Minami:2020odp, 2022PhRvL.128i1302D, Eskilt:2022wav, 2022PhRvD.106f3503E, AtacamaCosmologyTelescope:2025blo, 2025arXiv250913654D}. We further justify the validity of this prior in \cref{discussion_on_priors}. Lastly, $\alpha$ is sampled from a Gaussian distribution $\alpha \sim \mathcal{N}(0,0.5\degree)$, since the typical systematic uncertainty on polarization angle calibration is currently $\approx0.5\degree$, for experiments without an external absolute polarization calibrator~\cite{2010A&A...520A..13R, Planck:2016soo, 2020A&A...634A.100A}.

To account for the uncertainty in the power spectrum estimation, we generate a statistical ensemble of spherical harmonics coefficients of the observed $E$ and $B$ modes, $a_{\ell m}^E$ and $a_{\ell m}^B$, using the \texttt{hp.synalm} routine of \texttt{healpy}\footnote{\url{https://healpix.sourceforge.io}}~\cite{Gorski:2004by,Zonca:2019vzt} and \cref{miscal}. We smooth the observed $E$ and $B$ modes with a Gaussian beam of 30\,arcmin full-width at half maximum. We also generate Gaussian white noise realizations of $2\,\mu$K-arcmin, corresponding to the typical combined sensitivity of \textit{LiteBIRD}~\cite{2023PTEP.2023d2F01L}. We combine each single CMB and noise realization, and compute the corresponding angular power spectrum up to a maximum multipole of $\ell_{\mathrm{max}} = 512$:
\begin{equation}
\label{eq:estimated_spectrum}
\widehat{C}_\ell^{EB} = \frac{1}{2\ell + 1} \sum_{m = -\ell}^{\ell} \mathrm{Re}\left(a_{\ell m}^E a_{\ell m}^{B*}\right) \, .
\end{equation}
Since SBI methods become computationally challenging as the size of the data vector increases significantly, we do not retain all multipoles from $\ell = 2$ to $512$ of $\widehat{C}_{\ell}^{EB}$ to build our training set. 
To keep the data vector size manageable while preserving the full sensitivity to the reionization bump, we use all multipoles from $\ell =2$ to $30$ and then group higher multipoles from $31$ to $200$ into bins of width $\Delta \ell = 10$.

In this paper, we do not consider additional sources of uncertainty other than white noise, such as Galactic foregrounds or partial sky coverage. Therefore, $\widehat{C}_\ell^{EB}$ is solely affected by cosmic and noise variance. We expect SBI to be able to learn and marginalize over such observational uncertainties during inference. We leave the inclusion of these extra levels of complexity to future work.

\section{Results}\label{Resultats}

\subsection{Non-Gaussianity of $EB$ statistics}\label{Likelihood_eb}

In this analysis, we constrain $m_{\phi}$, $g\phi_{\mathrm{in}}/2$, and $\alpha$, exploiting only the information contained in the $EB$ power spectrum rather than the full set, $EE$, $BB$, and $EB$, of polarization spectra. As shown in \cref{eq:estimated_spectrum}, the computation of $\widehat{C}_{\ell}^{EB}$ involves the product of two correlated Gaussian random fields, $a_{\ell m}^{E}$ and $a_{\ell m}^{B}$. Consequently, the likelihood $p(\widehat{C}_{\ell}^{EB}|{C}_{\ell}^{EB})$, describing the probability of the estimated $\widehat{C}_{\ell}^{EB}$ given a fiducial power spectrum ${C}_{\ell}^{EB}$, is non-Gaussian and analytically intractable at low multipoles  ($\ell \lesssim 10$),  even in the ideal case of full-sky and noiseless observations \cite{2006MNRAS.372.1104P,Hamimeche:2008ai,Gerbino:2019okg}. At higher multipoles, the central limit theorem makes the likelihood distribution tend toward a Gaussian because the power spectrum at a given multipole $\ell$ is estimated from $2\ell+1$ independent $m$ modes. In general, the distribution becomes analytically intractable when accounting for realistic complications such as partial sky coverage, complex instrumental noise, and foreground contamination. 

We exploit the capability of SBI, specifically the NLE approach, to estimate the true $p(\widehat{C}_{\ell}^{EB}|{C}_{\ell}^{EB})$ likelihood and highlight its non-Gaussian nature at low multipoles. We train the likelihood estimator on $N_\mathrm{sim}=10^5$ full-sky and CMB-only simulations of both the estimated ($\widehat{C}_\ell^{EB}$) and fiducial ($C_\ell^{EB}$) spectra over the multipole range $2 \leq \ell \leq 5$. The fiducial spectra are generated from parameters  $m_{\phi}$, and $g\phi_{\mathrm{in}}/2$ that are drawn from the prior. Due to the assumption of isotropy in our full-sky setup, the estimates from different multipole bins are uncorrelated. As a result, the distribution of $\widehat{C}_{\ell}^{EB}$  over the range $2\leq \ell \leq 5$  is fully determined by the fiducial values $C_{\ell}^{EB}$ within the same range.

\begin{figure}
    \centering
    \includegraphics[width=\textwidth]{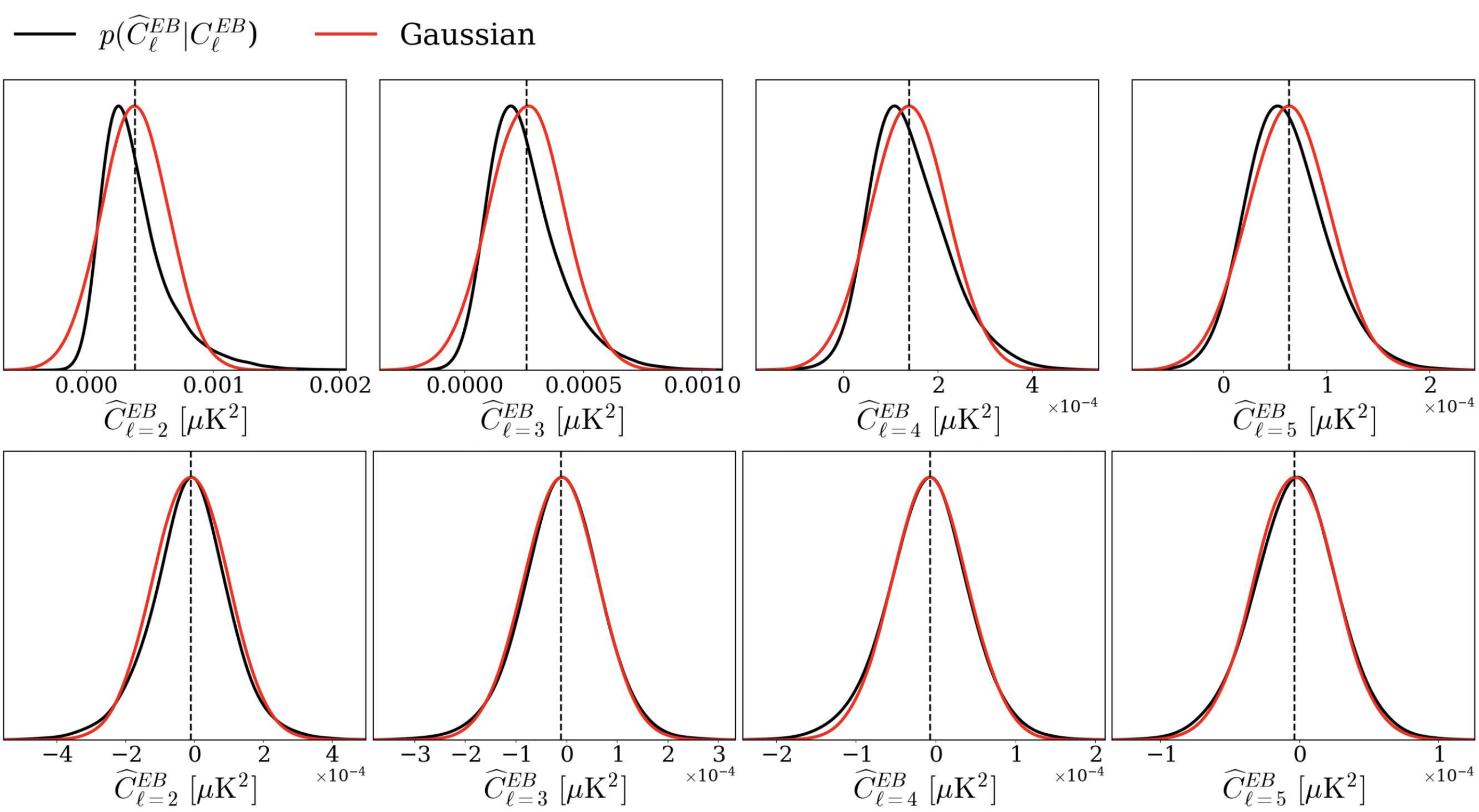} 
    \caption{Statistical distributions of the $EB$ power spectrum estimator $\widehat{C}_{\ell}^{EB}$, given a fiducial spectrum $C_{\ell}^{EB}$, for $\ell = 2, 3, 4, 5$. The distributions obtained with NLE (black) are compared to those assuming a Gaussian distribution (red). The upper panels correspond to a dark energy case with fiducial mass of $m_{\phi} = 10^{-32}$\,eV, and the lower panels to a dark matter case with $m_{\phi} = 10^{-29}$\,eV, for which the reionization $EB$ signal is nearly zero. The vertical dashed lines indicate the fiducial values of $C_{\ell}^{EB}$.  }\label{Cl-likelihood}
\end{figure}

    Once estimated, the likelihood $p(\widehat{C}_{\ell}^{EB}|C_{\ell}^{EB})$ is evaluated for two fiducial $C_{\ell}^{EB}$ that have different masses: one with $m_{\phi} = 10^{-32}$\,eV for which the $EB$ reionization bump is maximum (dark energy), and another with a larger mass for which the $EB$ signal nearly vanishes at low-$\ell$ (dark matter). Both fiducial spectra have the same $g \phi_{\mathrm{in}}/2 = 0.35^\circ$.  In \cref{Cl-likelihood}, the likelihood distributions obtained with NLE are shown for both the dark energy and dark matter scenarios and compared with a common Gaussian approximation of the likelihood, for which $\langle \widehat{C}_{\ell}^{EB} \rangle \approx C_{\ell}^{EB}$  and  $ \sigma^2(\widehat{C}_{\ell}^{EB}) = \langle (\widehat{C}_{\ell}^{EB})^2 \rangle - \langle \widehat{C}_{\ell}^{EB} \rangle^2 \approx [C_{\ell}^{EE} C_{\ell}^{BB} +  (C_{\ell}^{EB})^{2}]/(2\ell+1)$ (see \cref{eq:estimated_spectrum}). For $g\phi_{\mathrm{in}}/ 2 = 0.35^\circ$, $C_{\ell}^{EB}$ shows a relatively large positive reionization signal in the dark energy regime.  As a result,  the likelihood $p(\widehat{C}_{\ell}^{EB} | C_{\ell}^{EB})$ becomes positively skewed, with a tail extending towards large values of $\widehat{C}_{\ell}^{EB}$. For negative values of $g \phi_{\mathrm{in}}/ 2$, the likelihood instead has negative skewness. In contrast, since $C_{\ell}^{EB}$ is close to zero in the dark matter regime, its likelihood is centered on zero, although it still has non-Gaussian heavy tails. 

\subsection{Exploring phenomenology with amortized inference}
\label{am_inf}

In this section, we adopt the more flexible approach of amortized inference to explore the phenomenology of the problem. We show posteriors for test data that are generated assuming different masses in \cref{de vs dm}, the impact of lensed $B$ modes on the constraining power in \cref{impact_Lensing}, and quantify the confidence at which the full shape of $C_\ell^{EB}$ can reject the dark energy interpretation of cosmic birefringence in \cref{probability}. These results are computed using the NPE method (and MAF normalizing flow) applied to $N_\mathrm{sim}={10}^{5}$ simulations of $\widehat{C}_{\ell}^{EB}$ generated for various values of $\boldsymbol{\theta}=(m_{\phi}, g\phi_\mathrm{in}/2,\alpha)$ that are drawn from the prior. Details on the neural network architecture and posterior diagnostic tests are presented in \cref{NN-hp,coverage_tests}, respectively.

\subsubsection{Dark energy vs dark matter scenarios}\label{de vs dm}

\begin{figure}
    \centering
    \begin{subfigure}[t]{0.4\textwidth}
        \includegraphics[width=\linewidth]{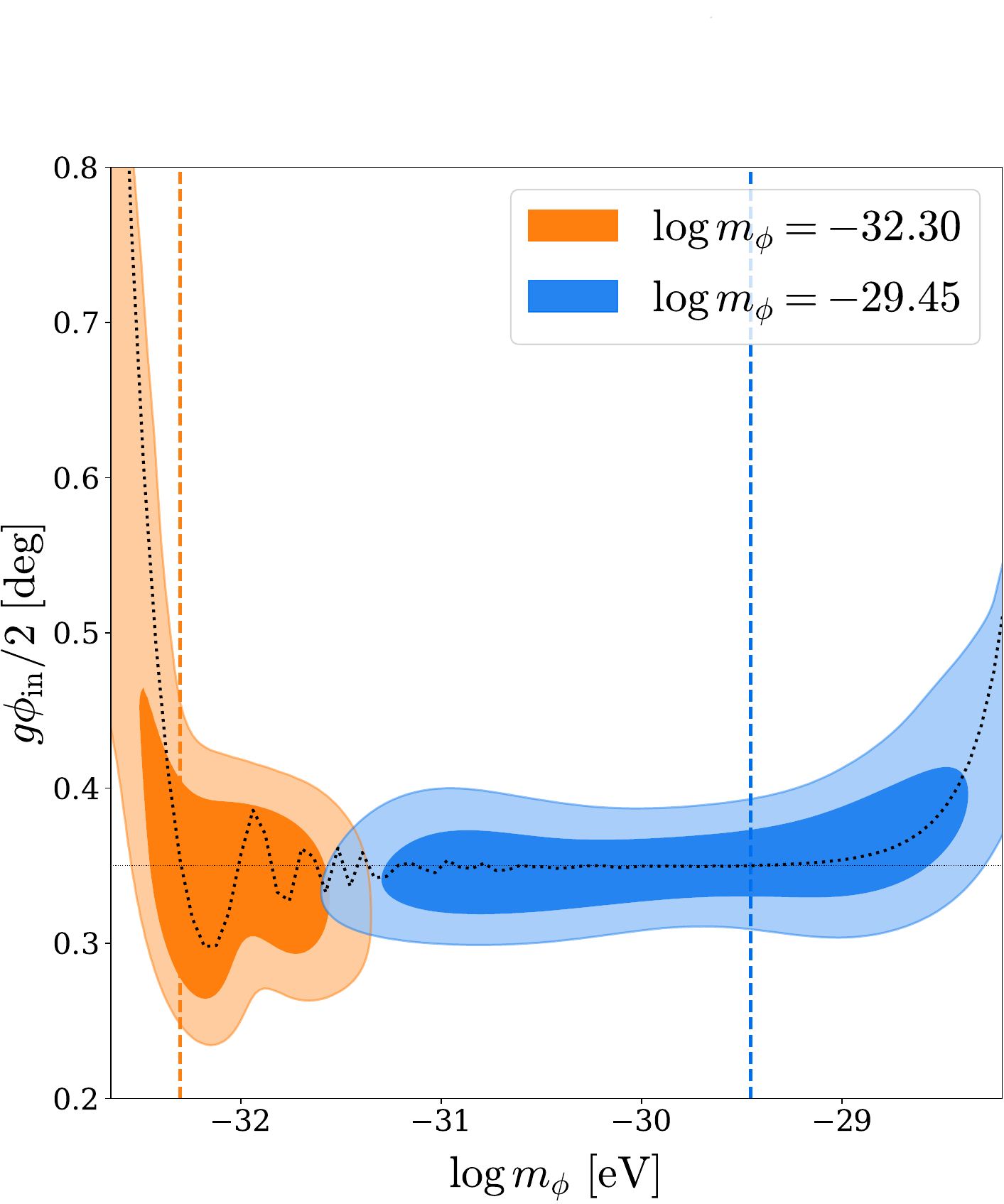}
    \end{subfigure}
    \hfill
    \begin{subfigure}[t]{0.55\textwidth}
        \includegraphics[width=\linewidth]{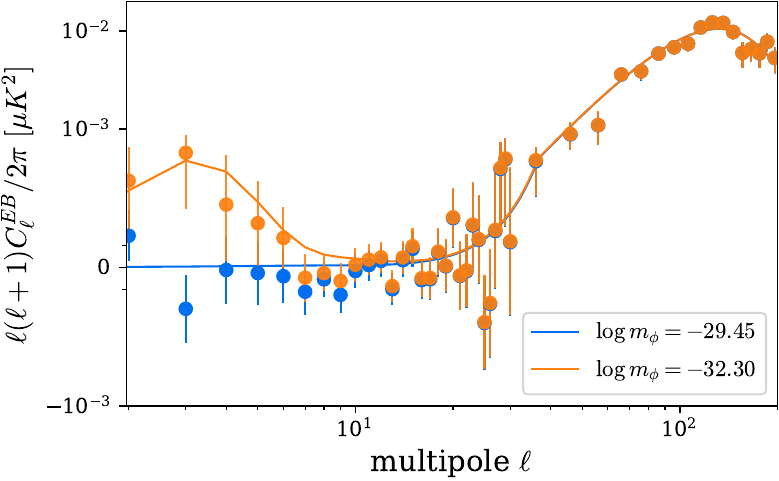}
    \end{subfigure}
    \caption{\textit{Left:} $68\%$ and $95\%$ highest-posterior density contours of $p(m_{\phi},g\phi_{\mathrm{in}}/2|\widehat{C}_{\ell}^{EB})$ for two fiducial masses: $\log m_{\phi}=-29.45$ (blue)  and $\log m_{\phi}=-32.30$ (orange). The black dotted line shows the required value of $g\phi_{\mathrm{in}}/2$ for the $EB$ power spectrum at high mutipoles to match the fiducial cosmic birefringence amplitude ($0.35^\circ$) at recombination. \textit{Right:} The corresponding $EB$ power spectra (same realization). Solid lines represent the fiducial $C_{\ell}^{EB}$ and the points show the estimated $\widehat{C}_{\ell}^{EB}$.
  Note the change from logarithmic to a linear scale for $\ell (\ell+1) C_{\ell}^{EB}/2\pi<5.10^{-4}\ \mu K^2$ to accommodate negative fluctuations.}
    \label{Contour_no_alpha}
\end{figure}

First, we explore the simplest case for which only $m_\phi$ and $g\phi_{in}/2$ change across simulations, fixing $\alpha =0 \degree$, and estimate $p(m_{\phi},g\phi_{\mathrm{in}}/2|\widehat{C}_{\ell}^{EB})$, as a validation test of our framework. In \cref{Contour_no_alpha} (left panel),  we show the expected $68\%$ and $95\%$ highest-posterior density (HPD) contours in the ($m_{\phi}$, $g\phi_{\mathrm{in}}/2$) plane, evaluated at $\widehat{C}_{\ell}^{EB}$ with $\log m_{\phi}^{\mathrm{fid}}=-32.30$ (orange) and $\log m_{\phi}^{\mathrm{fid}}=-29.45$ (blue). Both $EB$ power spectra (right panel) have fiducial  $g\phi_{\mathrm{in}}/2 = 0.35\degree$.

\begin{figure}
    \centering
    \includegraphics[width=0.85\textwidth]{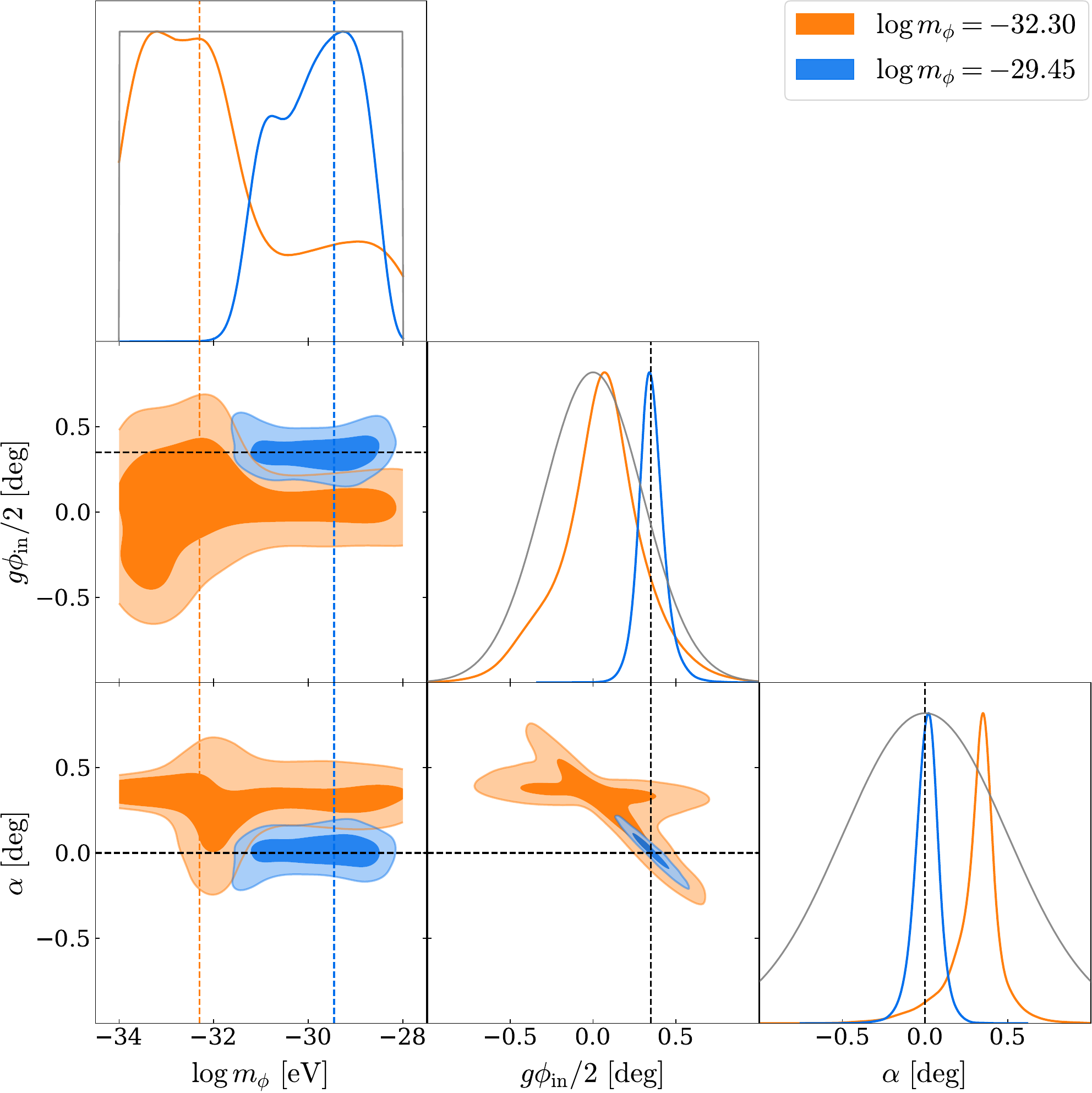} 
    \caption{Marginal posterior distributions of $m_{\phi}$, $g\phi_{\mathrm{in}}/2$, and $\alpha$. The black dashed lines represent the ground truth parameters: $\alpha = 0 \degree$  and $g\phi_{\mathrm{in}}/2=0.35\degree$. The blue and orange dashed lines show two fiducial masses: $\log m_{\phi}=-29.45$   and $\log m_{\phi}=-32.30$, respectively. The grey line represents the prior distribution used for each parameter.}\label{Posterior_amortized}
\end{figure}

For $\log m_\phi^\mathrm{fid} = -32.30$,  the data favor masses below $\sim 10^{-32}$\,eV, for which the predicted $EB$ power spectrum shape is similar to that of $EE$, therefore consistent with the fiducial model. In contrast, for $\log m_\phi^\mathrm{fid} = -29.45$, the reionization bump of $\widehat{C}_{\ell}^{EB}$ is suppressed, rendering masses below $\sim 10^{-32}$\,eV incompatible with the data. Since the $EB$ signal is weaker for ultralight ($m_{\phi} \ll 10^{-32}$\,eV) and heavier fields ($m_{\phi} \gtrsim 10^{-29}$\,eV) as explained in \cref{axion_dynamics}, the required value of $g \phi_{\mathrm{in}}/2$ must increase to reach the fiducial amplitude (black dotted line).\footnote{Following \cite{Nakatsuka_2022}, for each mass, we compute the cosmic birefringence rotation at recombination: $\beta^\mathrm{rec}(m_\phi) = \sum_{\ell=11}^{500} \beta_\ell^\mathrm{eff}/490$, with $\beta_\ell^\mathrm{eff}=\arcsin (2x_\ell/(1+x_\ell^2))/4$ and $x_\ell=C_\ell^{EB}/C_\ell^{EE}$. By construction, $\beta^\mathrm{rec}$ reaches its maximum for masses that do not suppress high-$\ell$ modes (see \cref{axion_potential}).
The black dotted line in \cref{Contour_no_alpha} represents the required value of $g\phi_{\mathrm{in}}/2$ to  match our fiducial amplitude of $0.35^\circ$, while accounting for the suppression of $\beta^{\mathrm{rec}}$ across the mass range such that $g\phi_{\mathrm{in}}/2 \cdot \beta^{\mathrm{rec}}(m_{\phi})=0.35^\circ \cdot \beta^{\mathrm{rec,max}}$.}

\Cref{Posterior_amortized} shows the posterior when the miscalibration angle $\alpha$ is allowed to vary as well. 
The posterior $p(m_{\phi}, g\phi_{\mathrm{in}}/2, \alpha | \widehat{C}_{\ell}^{EB})$ is evaluated for the two same $EB$ power spectra shown in \cref{Contour_no_alpha}. The fiducial value for these two test datasets is set to $\alpha =0^\circ$.

For $\log m_{\phi} = -32.30$ (orange), $g\phi_{\mathrm{in}}/2$ and $\alpha$ are degenerate \cite{Sherwin:2021vgb,Nakatsuka_2022}, so that the fiducial model can be reproduced by different combinations of these two parameters. In particular, ultralight fields are consistent with the data through a polarization rotation of $\alpha \simeq 0.35^\circ$, while $g\phi_{\rm in}/2$ remains unconstrained, since reproducing the fiducial amplitude would require values much larger than those supported by the prior. For masses near the fiducial value, $\alpha$ and $g\phi_{\mathrm {in}}/2$ are fully degenerate, and any combination satisfying $\alpha +g\phi_{\mathrm {in}}/2 \simeq 0.35^\circ$ reproduces the data equally well. Heavier masses can also match the fiducial model provided that $g\phi_{\mathrm {in}}/2\simeq0^\circ$ and $\alpha \simeq 0.35^\circ$.

For $\log m_{\phi}=-29.43$ (blue), the degeneracy between $\alpha$ and $g\phi_{\mathrm{in}}/2$ is partially broken since the reionization bump is supressed, while $\alpha$ yields $C_\ell^{EB}\propto C_\ell^{EE}-C_\ell^{BB}$. For this reason, the data disfavor models for which the $EB$  reionization bump closely follows the shape of $C_{\ell}^{EE}$, restricting the posterior to masses above $10^{-32}$\,eV. We emphasize that, given our current configuration, it is not possible to distinguish between masses below $10^{-32}$\,eV, as the corresponding low-$\ell$ $EB$ signal is too low and becomes dominated by cosmic variance and instrumental noise (see \cref{Contour_no_alpha}). These results highlight the importance of using the full shape of the $EB$ power spectrum, and especially the largest angular scales, to probe the nature of the scalar field.

\subsubsection{Impact of lensing $B$ modes}\label{impact_Lensing}

As discussed in \cref{Likelihood_eb}, the variance of $\widehat{C}_{\ell}^{EB}$ scales as $\propto C_{\ell}^{EE}C_{\ell}^{BB}$. Therefore, lensing $B$ modes contribute significantly to the variance of the estimated $EB$ spectrum. Although lensing dominates the $B$-mode power at intermediate and high multipoles, its contribution is also not negligible at low multipoles, particularly in the absence of primordial tensor modes and for weak $EB$ signals. Several analyses~\cite{2025JCAP...07..009L, Naokawa:2023upt, BICEPKeck:2024cmk} demonstrated that lensing $B$ modes lead to larger uncertainties on cosmic birefringence constraints.

\begin{figure}
    \centering
    \includegraphics[width=\textwidth]{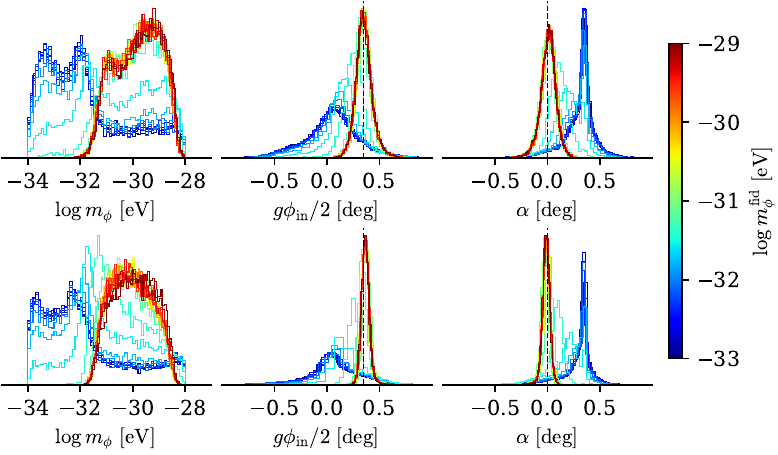} 
    \caption{Posterior distributions on $m_{\phi}$, $g\phi_{\mathrm{in}}/2$, and $\alpha$ with different fiducial masses (indicated with the colorbar). Each posterior is conditioned on the same CMB and noise realization, for both upper and lower panels. The black dashed lines represent $g\phi_{\mathrm{in}}/2 = 0.35 \degree$ and $\alpha =0\degree$. The upper panels show the case in which simulations include lensing $B$ modes, while the lower panels do not include them. The upper and lower panels  are normalized such that the area under each histogram is 1. }\label{Posterior_lensed_unlensed}
\end{figure}

We produce a new set of $N_{\mathrm{sim}} = {10}^{5}$ $\widehat{C}_{\ell}^{EB}$ simulations without lensing, and estimate the joint posterior distribution $p(m_{\phi},g\phi_{\mathrm{in}}/2,\alpha|\widehat{C}_{\ell}^{EB})$ with NPE. Even in the absence of lensing or primordial tensor modes, there remains a small $BB$ signal due to birefringence (see \cref{miscal}). In \cref{Posterior_lensed_unlensed}, we evaluate the posterior distributions for different values of $m_{\phi}$ ranging from $10^{-33}$ to ${10}^{-29}$\,eV. The upper panels correspond to the case with the lensing $B$ modes, while the lower panels show the case without them.
As explained in \cref{axion_dynamics}, for a fixed $g\phi_{\mathrm{in}}/2$, the overall amplitude of $C_{\ell}^{EB}$ varies across $m_{\phi}$. Since observational constraints at $\ell >50$ suggest $\beta$ close to $0.35^\circ$, we only use fiducial values of $g\phi_\mathrm{in}/2$ which ensure that our 50 test power spectra have a high-$\ell$ amplitude consistent with a uniform rotation of $0.35^\circ$. For each mass, we find the corresponding $g\phi_{\mathrm{in}}/2$ values by miniziming $\chi^2= \sum_{\ell>50} \big[C_{\ell}^{EB,\mathrm{ref}} - (g\phi_{\mathrm{in}}/2) C_{\ell}^{EB}(m_{\phi},g\phi_{\mathrm{in}}/2=1^\circ)\big]^2 / (\sigma_{\ell}^{EB})^2$, where $C_{\ell}^{EB,\mathrm{ref}}$ is a reference template of a uniform $0.35^\circ$ rotation across angular scales. In this way, our constraints would be solely driven by low-$\ell$ variations of $C_{\ell}^{EB}$.

As expected, the parameter estimation is more precise in the absence of lensing $B$ modes, especially for $\alpha$ and $g\phi_{\mathrm{in}}/2$. The lower variance of $\widehat{C}_{\ell}^{EB}$ at low multipoles helps break the degeneracy between $g\phi_\mathrm{in}/2$ and $\alpha$ by improving the sensitivity to the reionization bump.
More importantly, the absence of lensing enhances the difference between the dark energy- and dark matter-like posterior distributions of $m_{\phi}$. While dark matter posteriors ($m_\phi\gtrsim 10^{-31}$) show a smooth decrease in probability towards lower masses in the upper panels, we see a sharper cutoff at $m_\phi=10^{-32}$\,eV in the lower panels. Therefore, delensing could play an important role in determining the origin of cosmic birefringence.

\subsubsection{Excluding the dark energy interpretation}\label{probability}

Throughout \cref{am_inf}, we demonstrated that the constraints on $m_\phi$ are highly sensitive to the full shape of the $EB$ power spectrum, offering a distinctive probe on the nature of $\phi$. In this section, we quantify how the posterior on $m_{\phi}$ shifts, when conditioned on $EB$ power spectra generated with different fiducial masses. We use the probability $p(m_{\phi} < 10^{-32} \ \mathrm{eV})$ as a metric to measure the confidence level with which we would reject the dark energy interpretation of the observed cosmic birefringence signal, if a dark matter axion field sources it.

Similarly to \cref{Posterior_lensed_unlensed}, we generate $EB$ power spectra for $N$ fiducial masses and evaluate the posterior on $m_{\phi}$, conditioned at each simulation. This results in $N$ posterior distributions on $m_{\phi}$ from which we compute  $p(m_{\phi} < 10^{-32} \ \mathrm{eV})$, and assess how this probability evolves under different fiducial values of $m_{\phi}$. 

\begin{figure}
    \centering
    \includegraphics[width=0.90\textwidth]{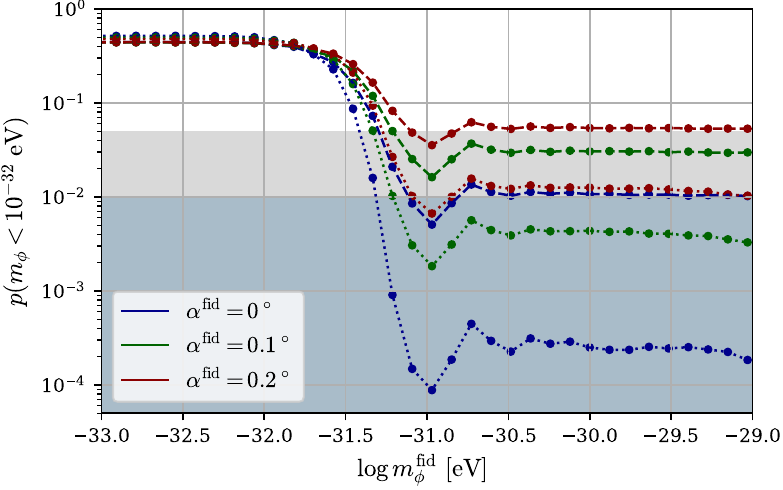} 
    \caption{Probability of $m_{\phi}<10^{-32}$\,eV for different fiducial masses $m_{\phi}^{\mathrm{fid}}$ (a single point represents the averaged probability over 50 realizations of CMB and noise). The dashed and dotted lines represent the case with and without lensing $B$ modes, respectively. The probability is shown for three fiducial values of $\alpha = 0^\circ$ (blue), $ 0.1^\circ$ (green), and $0.2^\circ$ (red). The shaded areas highlight the $<1\%$ (blue) and $<5\%$ (grey) probability regions. 
}\label{pb_plot}
\end{figure}

The probability $p(m_\phi<10^{-32}\,\mathrm{eV})$ is shown in \cref{pb_plot} for $\alpha^{\mathrm{fid}} =0^\circ$ (blue), $0.1^\circ$ (green), and $0.2^\circ$ (red), while the fiducial value of $g \phi_{\mathrm{in}}/2$  is computed for each mass so that the $EB$ power spectrum at high multipoles remains unchanged, as explained in \cref{impact_Lensing}. To mitigate the impact of cosmic variance and noise on this forecast, the probability is averaged over 50 independent simulations with different CMB and noise realizations.
For fiducial masses $m_\phi^\mathrm{fid}<10^{-32}$\,eV, the probability saturates to a constant $ p(m_\phi<10^{-32}\,\mathrm{eV} )\approx 40 - 50\,\%$ value as the $EB$ power spectrum is identical for all masses below $10^{-32}$\,eV. Going to heavier masses, all curves exhibit a decrease in probability, reflecting the transition between the two posterior regimes  that occurs for masses around  $10^{-32}$ to $10^{-31}$\,eV, as shown in \cref{de vs dm}. This result is consistent with what was found in \cite{Nakatsuka_2022}. 
As anticipated from \cref{Posterior_lensed_unlensed}, this drop in probability is even sharper when lensing $B$ modes are removed (solid vs dashed lines). Without lensing $B$ modes, the scale-dependent signal is easier to distinguish, thus tightening the constraints. For $m_{\phi}^{\mathrm{fid}} > 10^{-31}$\,eV, the probability appears to level off near a minimum value. Such large fiducial masses induce a sufficient scale-dependent effect to confidently exclude the dark energy origin of the signal.

Overall, \cref{pb_plot} highlights that delensing enables us to reject a dark energy scenario for $m_{\phi}^{\mathrm{fid}} \gtrsim 10^{-31.3}$\,eV at the $99\%$ confidence level, even in the presence of miscalibrations up to $0.2^\circ$. In the presence of lensing, $p(m_{\phi} < 10^{-32} \ \mathrm{eV})$ stays at the level of $1\%$ when $\alpha^{\mathrm{fid}} = 0^\circ$, but increases up to $5\%$ for larger miscalibrations.
Therefore, we conclude that $\alpha$ miscalibrations have a significant impact on the probability of rejecting a dark energy interpretation.

\subsection{Towards real data analysis: sequential inference}\label{SNPE}

The amortized inference approach of \cref{am_inf} required the generation of training sets with $N_\mathrm{sim}=10^5$ simulations. For a realistic analysis, simulations must include instrumental noise and systematic effects, the cleaning of Galactic foreground emission, and the correction of mode couplings and leakages introduced by masking. The computational cost to produce such a large number of high-fidelity simulations is prohibitive. Furthermore, since real CMB measurements consist of a single observed power spectrum, the generalization of the posterior model over multiple observational data provided by the amortized inference is unnecessary. These two arguments motivate the SNPE approach, which is expected to achieve convergence to the true posterior with fewer simulations by adaptively focusing the inference process on regions of high posterior probability. 

Here, we employ truncated sequential neural posterior estimation (TSNPE)~\cite{2022arXiv221004815D}, which uses the inferred posterior from each round to identify the highest probability regions (HPR) of the parameter space. TSNPE filters out prior regions that fall outside the HPR defined by $1-\epsilon$, where $\epsilon$ is a rejection threshold. This leads to a truncated version of the prior, which is then used as the proposal distribution for the next round. The rejection threshold $\epsilon$ essentially indicates the fraction of low-probability regions that are excluded from the prior. Therefore, this parameter is crucial since,  if too large, some high-density regions might be excluded, potentially biasing the posterior. However, \cite{2022arXiv221004815D} demonstrated that for $\epsilon \le 10^{-4}$, the HPR retains nearly $100\%$ of the true posterior samples, confirming that our chosen value of $\epsilon$ is sufficiently small. 

\begin{figure}[t]
    \centering
    \includegraphics[width=0.85\textwidth]{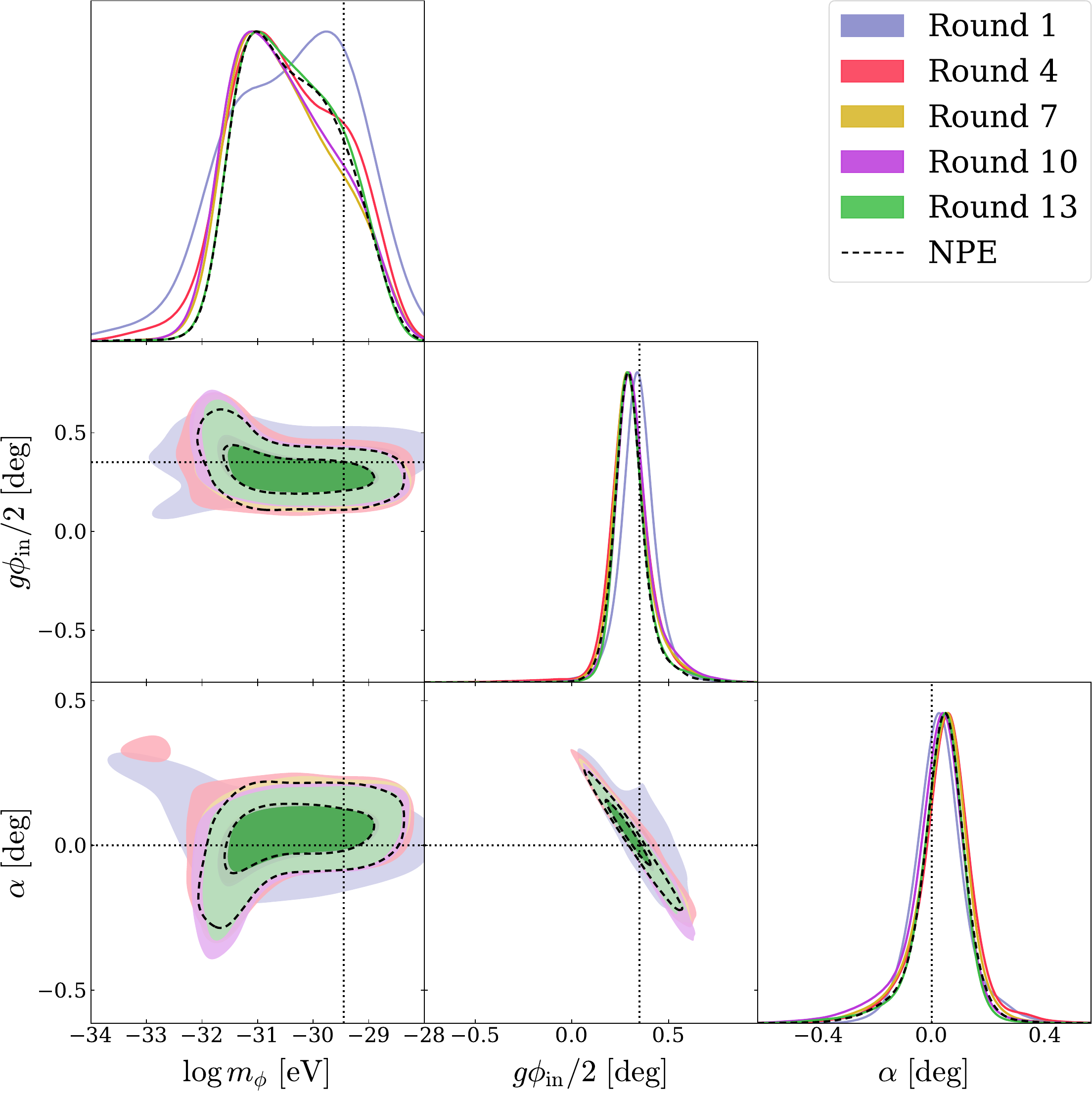} 
    \caption{Marginal posterior distributions for $m_{\phi}$, $g\phi_{\mathrm{in}}/2$, and $\alpha$ for several TSNPE rounds. The posterior distribution obtained with the amortized NPE ($N_\mathrm{sim}=10^5$ simulations) from \cref{am_inf} is shown with the black dashed contours. The black dotted lines show the true parameters.}\label{posterior_SNPE}
\end{figure}

We opt for a TSNPE configuration with 15 rounds, allocating more simulations to the initial rounds and gradually reducing the number of simulations in later rounds, for a total of 30000 simulations. This configuration accommodates the high variability of posteriors with fiducial parameters (see \cref{Posterior_amortized}), with earlier rounds capturing the global structure and later rounds refining it locally. The rejection threshold is $\epsilon = {10}^{-4}$, meaning that we reject $0.01\,\%$ of the prior samples that are outside the HPR. In each round, the posterior is evaluated for a specific $\widehat{C}_{\ell}^{EB}$ with $\alpha = 0^\circ$ and $g\phi_{\mathrm{in}}/2= 0.35^\circ$.

The colored contours in \cref{posterior_SNPE} show that the TSNPE posterior for $\alpha$ and $g\phi_{\mathrm{in}}/2$ nearly converge to the NPE reference (obtained with $N_\mathrm{sim}=10^5$ simulations, black dashed line), already at round 4. Subsequent rounds (7, 10, and 13) adjust the distribution tails via truncation, resulting in consistent contours with the amortized NPE solution. The constraints on $m_{\phi}$, however, evolve considerably even in the latest inference rounds, closely matching the amortized inference posterior at round 13. This suggests that our simulations are less sensitive to $m_{\phi}$, making it a more challenging parameter to infer than $g\phi_{\mathrm{in}}/2$ and $\alpha$. For clarity, we do not show the posterior contours of rounds 14 and 15 in \cref{posterior_SNPE}, since TSNPE has already converged at round 13.

Minor differences persist between TSNPE and NPE solutions, likely due to the use of an emulator in TSNPE rather than the exact \texttt{birefCLASS} spectra in NPE. Overall, 13 rounds of TSNPE are enough to closely reproduce the NPE reference using only 28600 simulations. This constitutes a reduction of the simulation budget by a factor of $\approx 3.5$.

\section{Discussion}\label{Discussion}

In this paper, we demonstrated how to use SBI to constrain the scalar field mass ($m_{\phi}$), the cosmic birefringence amplitude ($g \phi_{\mathrm{in}}/2$), and the polarization angle miscalibration ($\alpha$), from the CMB $EB$ power spectrum. The intractability of the low-$\ell$ $EB$ likelihood makes SBI a natural choice, as it performs parameter inference without relying on an explicit likelihood function. In \cref{Likelihood_eb}, we demonstrated the non-Gaussian nature of this likelihood by applying NLE to infer the distribution of the estimated $EB$ power spectrum, $\widehat{C}_{\ell}^{EB}$, given a fiducial power spectrum, $C_{\ell}^{EB}$, at $\ell = 2,3,4,5$. In a dark energy-like scenario, characterized by an enhanced reionization bump,  the likelihood exhibits a strong departure from Gaussianity and is positively skewed (assuming a positive amplitude). Conversely, in a dark matter scenario, the reionization bump nearly vanishes and the likelihood takes the form of a symmetric non-Gaussian distribution, with pronounced heavy tails. 

In \cref{de vs dm}, we used the NPE pipeline with an amortized inference approach, to estimate the posterior distribution $p(\log m_{\phi},g \phi_{\mathrm{in}}/2,\alpha|\widehat{C}_{\ell}^{EB})$. The posterior was then evaluated for two realizations of the $EB$ power spectrum: a dark energy case ($\log m_{\phi} = -32.30$) and a dark matter case ($\log m_{\phi} = -29.45$). In the dark energy case, $g \phi_{\mathrm{in}}/2$ and $\alpha$ are fully degenerate. As a result, the full mass range is supported by the data, but with different combinations of $\alpha$ and $g\phi_{\mathrm{in}}/2$. In the dark matter case, the degeneracy between $g \phi_{\mathrm{in}}/2$ and $\alpha$ is partly broken since the cosmic birefringence effect affects differently the reionization scales ($\ell \lesssim 10$) and higher multipoles, while $\alpha$ uniformly rotates all angular scales. The resulting posterior on $m_{\phi}$ prefers $m_{\phi}>10^{-32}$ eV, reflecting the incompatibility of smaller mass with a suppressed reionization bump. Overall, our results show that the posterior alternates between two distinct regimes depending on whether the fiducial mass lies above or below $10^{-32}$ eV. This is consistent with \cite{Nakatsuka_2022}. 

In \cref{impact_Lensing}, we showed that the absence of lensed $B$ modes results in significantly tighter parameter constraints. This behavior is expected as the lensing $B$ modes contribute significantly to the overall variance of $\widehat{C}_{\ell}^{EB}$ \cite{2025JCAP...07..009L,Naokawa:2023upt}. This improvement suggests that forthcoming CMB experiments could benefit from incorporating delensing strategies for cosmic birefringence analyses \cite{Namikawa:2021gyh,LiteBIRD:2023aov}. The improvement provided by delensing is further illustrated in \cref{probability}, where we find that the dark energy interpretation of cosmic birefringence can be ruled out at the $99\%$ confidence level, for polarization angle miscalibrations up to $0.2^\circ$. In the lensed case, the constraint is less stringent, if the assumed miscalibration is $\gtrsim 0.1^\circ$. Overall, these results indicate that the combined effect of lensing $B$-mode variance and a non-zero polarization angle miscalibration reduces, but does not negate, our ability to clearly distinguish between dark energy and dark matter scenarios.

We considered an idealized cosmic variance-limited scenario, neglecting important sources of uncertainty present in real CMB analyses, i.e., Galactic foregrounds, masking, and instrumental systematic effects. Our primary goal was to present an alternative parameter inference approach that can handle non-Gaussianity of the $EB$ likelihood as discussed in \cref{Likelihood_eb}, even in this simplified setup.
Although these sources of uncertainty will inevitably degrade the signal-to-noise ratio and introduce correlations between multipoles, SBI is expected to recover the relevant features of the data, including nuisance parameters, given a sufficient amount of simulations and a well-optimized neural network. This expectation, of course, relies on our ability to accurately simulate the effect of all revelant sources of uncertainty. In view of a real data analysis, for which high-fidelity and experiment-specific simulations are limited, we used the sequential inference approach of \cref{SNPE} as a first attempt to reduce the simulation budget. By tailoring the posterior estimate to each observation and adjusting the proposal distribution across successive rounds, this method allowed us to reduce the simulation budget by a factor of $\approx 3$, with respect to an amortized inference approach (\cref{am_inf}). While this is not a dramatic improvement, additional data-compression techniques \cite{Alsing:2018eau,Jeffrey:2020xve,Dirmeier:2023sni} could be used to further reduce the required number of simulations. As noted in \cref{sims}, a CMB power spectrum emulator \cite{2022MNRAS.511.1771S} offers a fast alternative to the usual Boltzmann solvers, a useful feature given the need to produce substantial simulation datasets for SBI. Further implementations capable of emulating CMB systematic effects \cite{Campeti:2025opc} could also be relevant for this study.

Throughout this paper, we used the power spectrum as a summary statistic, assuming a specific binning scheme (\cref{sims}) to keep the dimensionality of the data vector manageable. This choice is reasonable for the current analysis because we focus on the cosmic birefringence effect induced by very light fields, $m_{\phi} \lesssim 10^{-28}$\,eV, whose impact on the $EB$ power spectrum is mainly focused at reionization scales. Although it is still possible to include higher multipoles and explore a wider range of masses using SBI, one of the aforementioned data compression techniques would be required to keep the size of the data vector reasonable. Alternatively, one could adopt a standard likelihood approach for the high multipole regime and combine it with the low multipole likelihood obtained via NLE in section~\ref{Likelihood_eb}. 

As explained in \cref{sims}, we adopted a Gaussian prior for the cosmic birefringence amplitude $g \phi_{\mathrm{in}}/2$ centered at $0$ (corresponding to the expected $\Lambda \mathrm{CDM}$ value) with a standard deviation of $0.3^\circ$ (consistent with current observational limits on the cosmic birefringence angle at $\ell>50$~\cite{Minami:2020odp, 2022PhRvL.128i1302D,Eskilt:2022wav, 2022PhRvD.106f3503E, 2025arXiv250913654D, AtacamaCosmologyTelescope:2025blo}). In \cref{discussion_on_priors}, we assessed the impact of a less informative prior on $g \phi_{\mathrm{in}}/2$ (flat prior between $-1^\circ$ and $1^\circ$). We find that this change affects the results in the dark energy regime, because $\alpha$ and $g\phi_{\mathrm{in}}/2$ are degenerate, making the constraints prior-driven (otherwise any $\alpha + g\phi_{\mathrm{in}}/2$ would reproduce the $\widehat{C}_{\ell}^{EB}$ amplitude). However, the impact remains limited, since the narrower prior on $\alpha$ continues to dominate. This result highlights the importance of a precise polarization angle calibration. In the dark matter regime, the constraints are weakly sensitive to the prior on $g\phi_{\mathrm{in}}/2$ and are essentially driven by the data.

A natural extension of this work would be to assume a \textit{LiteBIRD} \cite{2023PTEP.2023d2F01L} instrumental configuration, which will measure the large angular scales of the polarization CMB power spectra, the most sensitive to ultralight fields. In this framework, the final parameter constraints would also depend on the fidelity of the $E$  and $B$ mode reconstruction, the efficiency of foreground cleaning, and the control of instrumental systematics.

\acknowledgments
We thank K. Murai, T. Namikawa, and F. Naokawa for their help with \texttt{birefCLASS}, and N. Anau Montel for useful discussions on the SBI. This analysis benefited from the following packages:  \texttt{numpy} \cite{Harris:2020xlr}, \texttt{matplotlib} \cite{Hunter:2007ouj}, \texttt{healpy} \cite{Gorski:2004by,Zonca:2019vzt}, \texttt{sbi} \cite{Tejero-Cantero:2020sbi, Boelts:2025sbiReloaded}, \texttt{getdist} \cite{Lewis:2019xzd}, \texttt{CosmoPower} \cite{2022MNRAS.511.1771S} and \texttt{optuna} \cite{2019arXiv190710902A}. This work was supported in part by the Excellence Cluster ORIGINS which is funded by the Deutsche Forschungsgemeinschaft (DFG, German Research Foundation) under Germany’s Excellence Strategy: Grant No.~EXC-2094 - 390783311. This work has also received funding from the European Union's Horizon 2020 research and innovation programme under the Marie Skłodowska-Curie grant agreement no.\ 101007633. The Kavli IPMU is supported by World Premier International Research Center Initiative (WPI), MEXT, Japan. Authors acknowledge partial support by the Italian Space Agency LiteBIRD Project (ASI Grants No. 2020-9-HH.0 and 2016-24-H.1-2018), and the Italian Space Agency Euclid Project, as well as the InDark and LiteBIRD Initiative of the National Institute for Nuclear Phyiscs, and the RadioForegroundsPlus Project HORIZON-CL4-2023-SPACE-01, GA 101135036, and Project SPACE-IT-UP  by the Italian Space Agency and Ministry of University and Research, Contract Number  2024-5-E.0.

\appendix

\section{Power spectrum emulator}\label{Emulator_validation}

In the sequential inference analysis presented in \cref{SNPE}, we replace \texttt{birefCLASS} \cite{Nakatsuka_2022,Murai:2022zur,Naokawa:2023upt} with a power spectrum emulator, which significantly accelerates the generation of simulations. The emulator is designed to reproduce theoretical power spectra ($BB$, $EB$) as a function of $m_{\phi}$ and $g \phi_{\mathrm{in}}/2$. As discussed in \cref{axion_dynamics}, the  $EB$ power spectrum follows a similar shape for $m_\phi<10^{-32}$\,eV, but changes noticeably at low mutipoles for larger $m_{\phi}$, with a suppression of the reionization bump. Hence, to improve the accuracy of emulated power spectra, we partition the training data for $m_{\phi} < 10^{-32}$\,eV and $m_{\phi} > 10^{-32}$\,eV and train two separate emulators.  During sequential inference, the appropriate emulator is selected depending on whether the mass sampled from the proposal distribution is above or below the $10^{-32}$\,eV threshold. 

We use $N_\mathrm{sim}=10^5$ $EB$ power spectra calculated for various values of $m_{\phi}$ and $g\phi_{\mathrm{in}}/2$ up to $\ell_{\mathrm{max}} = 600$ (same theoretical power spectra than in \cref{de vs dm}). Since multipoles are uncorrelated, higher multipoles are not necessary for computing $\widehat{C}_{\ell}^{EB}$ with $\ell_{\mathrm{max}}=512$ (see \cref{sims}).  Rather than training the emulator directly on these power spectra, we first apply principal component analysis (PCA)~\cite{Jolliffe:2002}  to reduce the dimensionality of the data, using $128$ PCA components. This transformation allows the emulator to focus on the most informative features of the spectra, improving the training efficiency. 

We validate the emulator by comparing the $EB$ spectra it generates with the exact calculation from \texttt{birefCLASS}.  In particular, we emulate $EB$ power spectra for a test set of 500 simulations and compute the ratio of the residuals compared to \texttt{birefCLASS}, over the total expected uncertainty:
\begin{equation}\label{delta_cl}
    \frac{\Delta C_\ell^{EB}}{\sigma_\ell^{EB}}  = \frac{|C_{\ell}^{EB,\mathrm{em}} - C_{\ell}^{EB,\mathrm{exact}}|}{\sigma_{\ell}^{EB}}.
\end{equation}
Here, $\sigma_{\ell}^{EB}$ corresponds to the standard deviation per multipole, which includes cosmic variance and instrumental noise \cite{2022MNRAS.511.1771S}: 
\begin{equation}\label{sigma_ell_EB}
    \sigma_{\ell}^{EB} = \sqrt{\frac{(C_{\ell}^{EE} + N_{\ell})(C_{\ell}^{BB} + N_{\ell})+{C_{\ell}^{EB}}^{2}}{2\ell+1}},
\end{equation}
where $N_{\ell} = \big(2\mu \mathrm{K.arcmin} \frac{\pi}{10800}\big)^{2} \sim 3.38.10^{-7} \mu K^2\ \mathrm{str}$. Although the emulator is trained to reproduce theoretical power spectra, in \cref{sigma_ell_EB} we evaluate its accuracy compared to the total uncertainty arising from power spectrum estimation in simulations (\cref{eq:estimated_spectrum}). Indeed, the emulator will remain an acceptable approximation as long as it stays within the statistical uncertainty given by cosmic variance and instrumental noise. 

\begin{figure}
    \centering
    \includegraphics[width=\linewidth]{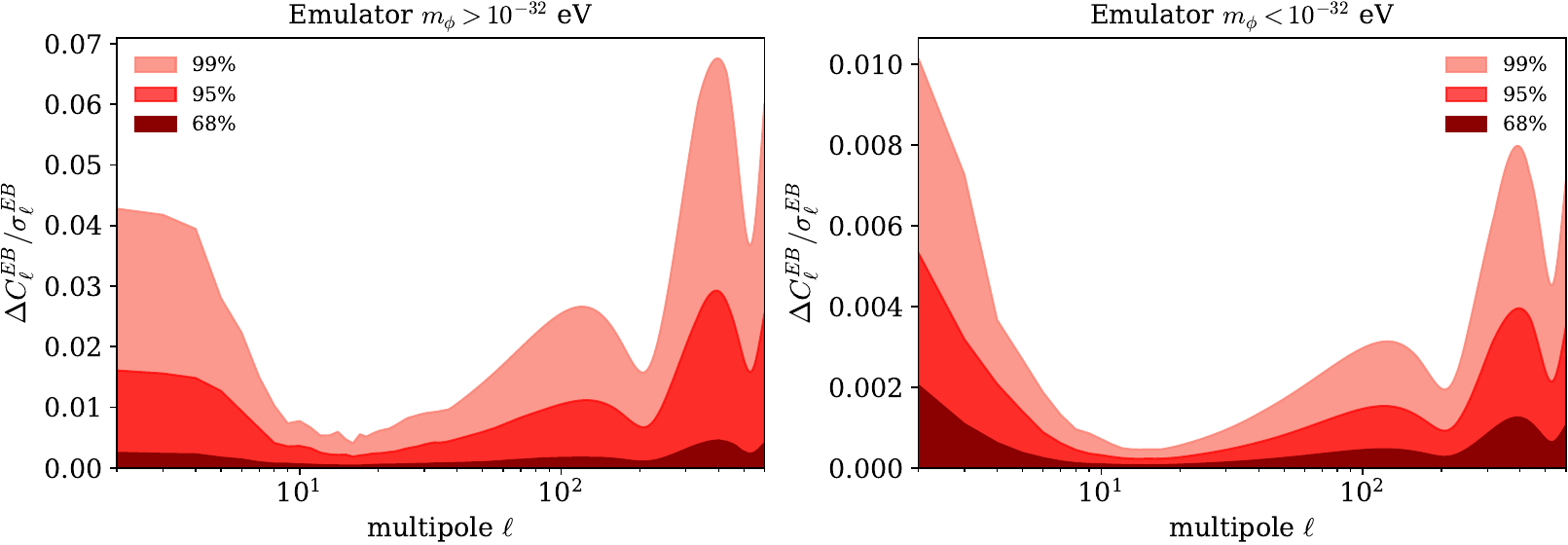}
    \caption{Relative error of the emulated $EB$ spectra with respect to the exact \texttt{birefCLASS} solution. 
    Shaded areas indicate the $68$, $95$, and $99\,\%$ percentiles of the relative error for a set of $500$ simulations. Left and right panels show the emulators for $m_{\phi}>10^{-32}$\,eV and  $m_{\phi}<10^{-32}$\,eV, respectively. }
    \label{emulator_validation_plot}
\end{figure}

\Cref{emulator_validation_plot} shows the relative error of the 500 emulated $EB$ power spectra. For $m_{\phi} > 10^{-32}$\,eV (left panel), the emulator error remains at the level of $4\%$ of the total statistical uncertainty for $99\%$ of the test dataset, at low multipoles ($\ell < 10$). At higher multipoles, $\Delta C_\ell^{EB}/\sigma_\ell^{EB}$ increases up to $7\%$ because ${\sigma}_\ell^{EB}$ grows smaller. Therefore, even minor differences between the emulated and exact power spectra lead to comparatively larger $\Delta C_\ell^{EB}/\sigma_\ell^{EB}$ values. For $m_\phi < 10^{-32}$\,eV (right panel), we notice a smaller error, with at most $\Delta C_\ell^{EB}/\sigma_\ell^{EB}  \approx 1\,\%$ at low multipoles. As only the overall amplitude of $EB$ changes in this mass range, the emulation of $C_\ell^{EB}$ is simpler than for the $m_\phi > 10^{-32}$\,eV case. While we only detailed the emulation procedure and validation tests for the $EB$ spectra here, analogous tests were applied to the $BB$ spectra. We do not apply the emulation procedure to the $EE$ power spectra, as they remain nearly unchanged under variations of $m_{\phi}$ and $g \phi_{\mathrm{in}}/2$.

\section{Impact of prior}\label{discussion_on_priors}

\begin{figure}
    \centering
    \includegraphics[width=0.90\textwidth]{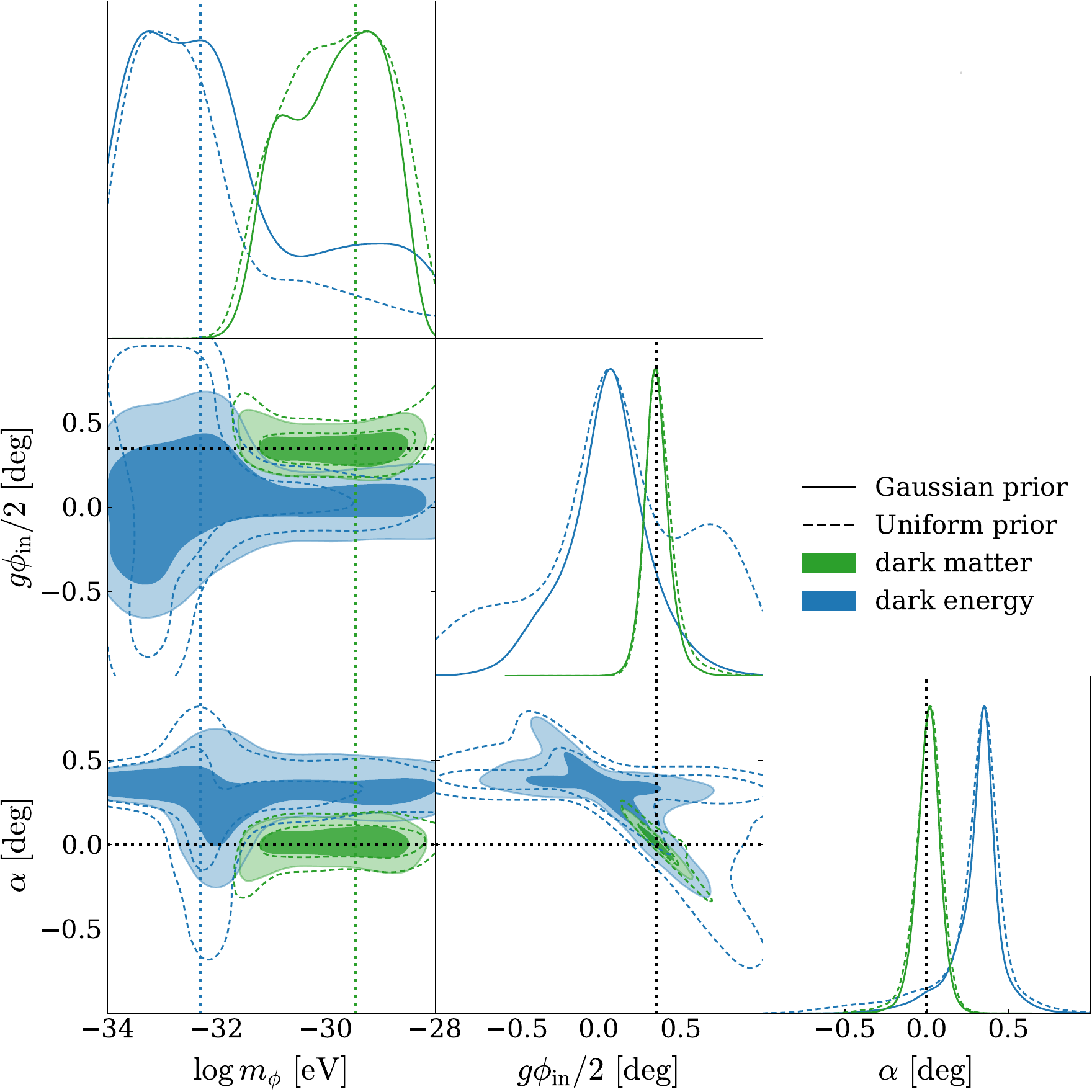} 
    \caption{Marginal posterior distributions of $m_{\phi}$, $g\phi_{\mathrm{in}}/2$, and $\alpha$, assuming two different priors for $g\phi_{\mathrm{in}}/2$: flat prior between $-1^\circ$ and $1^\circ$ (dashed) and a Gaussian prior centered at 0 with standard deviation of $0.3^\circ$ (solid). The blue and green contours correspond to the dark energy and dark matter scenarios, respectively. The black dotted lines represent the true parameters. }\label{constraints_diff_prior}
\end{figure}

We assess the impact of a less informative prior for the cosmic birefringence amplitude than that used in the main analysis, i.e., $g\phi_{\mathrm{in}}/2 \sim \mathcal{N}(0,0.3^\circ)$. We train the NPE (with identical configuration as in \cref{am_inf}) on $N_\mathrm{sim}=10^5$ simulated $EB$ power spectra that depend on $m_{\phi}$, $g\phi_{\mathrm{in}}/2$, and $\alpha$, now adopting a flat prior for $g\phi_{\mathrm{in}}/2$ between $-1^\circ$ and $1^\circ$. Because redefining the prior on $g \phi_{\mathrm{in}}/2$ requires generating a new set of theoretical power spectra, we use the emulator for fast simulation production. However, the emulator was trained with $g \phi_{\mathrm{in}}/2 \sim \mathcal{N}(0,0.3^\circ)$ and is expected to be reasonably accurate up to $3\sigma$. As larger $g \phi_{\mathrm{in}}/2$ values rely on interpolation, we can expect a small additional uncertainty in the emulator's prediction. The priors on $m_{\phi}$ and $\alpha$ remain unchanged from those specified in \cref{sims}. The posterior is then evaluated for the same $\widehat{C}_{\ell}^{EB}$ observations than in \cref{Posterior_amortized}. \Cref{constraints_diff_prior} shows the the marginal posterior for each parameter obtained assuming a flat prior (dashed line) and a Gaussian prior (solid line) for $g \phi_{\mathrm{in}}/2$. Blue (green) contours represent the dark energy (matter) scenario.

In the dark energy scenario, $\alpha$ and $g \phi_{\mathrm{in}}/2$ are degenerate; thus, priors are required to obtain closed confidence contours. Otherwise, there would be an infinite combination of $\alpha + g \phi_{\mathrm{in}}/2$ consistent with the observed $EB$ power spectrum amplitude. Consequently, the constraints are at risk of becoming prior-dominated. This is why we observe some impact when adopting a uniform prior on $g \phi_{\mathrm{in}}/2$ (mainly in the tails, and complementary for $\alpha$ and $g \phi_{\mathrm{in}}/2$), but it is not dramatic, since the narrower prior on $\alpha$ continues to control the constraints. This result highlights the importance of a precise polarization angle calibration, and demonstrates that adopting an informative prior in $g \phi_{\mathrm{in}}/2$ is valid, as long as the calibration of polarization angles remains reasonably accurate.
As mentioned in \cref{am_inf}, when $m_{\phi} \ll 10^{-32},\mathrm{eV}$, the values of $g\phi_{\mathrm{in}}/2$ supported by the prior are insufficient to match the fiducial amplitude. Widening the prior therefore allows a broader range of $g\phi_{\mathrm{in}}/2$ values, while $\alpha$ remains fixed at $\sim 0.35^\circ$.

In the dark matter scenario, the choice of the prior on $g \phi_{\mathrm{in}}/2$ has only a minor effect on the constraints since $\alpha$ and $g\phi_{\mathrm{in}}/2$ are now disentangled through the shape of the $EB$ power spectrum. In this case, constraints are primarily driven by the data.

\section{Neural network hyper-parameters}\label{NN-hp}

Optimization of neural network hyper-parameters involves tuning the elements that characterize the training process to minimize loss and improve generalization. During this process, both the training and validation losses are inspected to assess model performance. The final validation loss is used as the main objective metric, as it reflects how well the model is expected to perform on unseen data. Additionally, the gap between training and validation losses is measured as an indicator of overfitting: a small gap suggests good generalization, while a large gap indicates overfitting. We automate this process by using the \texttt{optuna}\footnote{\url{https://optuna.readthedocs.io/en/stable/index.html}} hyper-parameter optimization framework~\cite{2019arXiv190710902A}, which runs successive trials considering different combinations of hyper-parameters. For each trial, we evaluate the final validation loss, and select the best configurations that result in both the lowest validation loss and weak overfitting. The \texttt{optuna} framework is able to rank the hyper-parameters by importance, and to trace their individual impact on the validation loss. \Cref{hyperparams} presents the most relevant hyper-parameters, their functions, and their values.

\begin{table}
\centering
\begin{tabular}{|c|p{5cm}|c|c|}
\hline
Feature & Function & With lensing & Without lensing \\
\hline
\multirow{2}{*}{Learning rate} & Interval between neural network weight updates & \multirow{2}{*}{$5\cdot10^{-5}$} & \multirow{2}{*}{$2\cdot10^{-4}$} \\
\hline
\multirow{2}{*}{Batch size} & Number of simulations processed together & \multirow{2}{*}{256} & \multirow{2}{*}{128} \\
\hline
\multirow{2}{*}{Hidden features} & Number of units per neural network layer & \multirow{2}{*}{128} & \multirow{2}{*}{128} \\
\hline
\multirow{2}{*}{Number of transforms} & Number of transforms of the base distribution in the NF & \multirow{2}{*}{11} & \multirow{2}{*}{15} \\
\hline
\end{tabular}
\caption{Neural network hyper-parameters used in the NPE for simulations that include or exclude lensing $B$ modes.}
\label{hyperparams}
\end{table}

\section{Posterior checks and diagnosis}\label{coverage_tests}

Once the inference network is trained on a simulated data set, one may want to perform posterior checks and detect, if any, a bad posterior calibration. A straightforward approach is to first draw parameter values from the prior,
$\boldsymbol{\theta}_i \sim p(\boldsymbol{\theta})$, and then generate simulated data from the model,
$\boldsymbol{x}_i \sim p(\boldsymbol{x}|\boldsymbol{\theta}_i)$, using the simulator. For each simulated dataset $\boldsymbol{x}_i$, we compute the posterior distribution $p(\boldsymbol{\theta}|\boldsymbol{x}_i)$ and draw posterior samples  $\boldsymbol{\theta}_s^{(j)} \sim p(\boldsymbol{\theta}|\boldsymbol{x}_i)$. The so-called \textit{simulation-based calibration} (SBC)~\cite{Taltsetal.} relies on the principle that, if the posterior is well-calibrated, the posterior samples $\boldsymbol{\theta}_s^{(j)} $ should be distributed consistently with the prior. In practice, this can be assessed using rank statistics: for each simulation, one counts how many posterior samples $\boldsymbol{\theta}_s^{(j)}$ fall below the corresponding true parameter value $\boldsymbol{\theta}_i$. If the posterior is well-calibrated, the resulting ranks should be uniformly distributed across simulations. Deviations from uniformity indicate systematic biases or miscalibration in the inferred posterior.

\begin{figure}
    \centering
    \includegraphics[width=\textwidth]{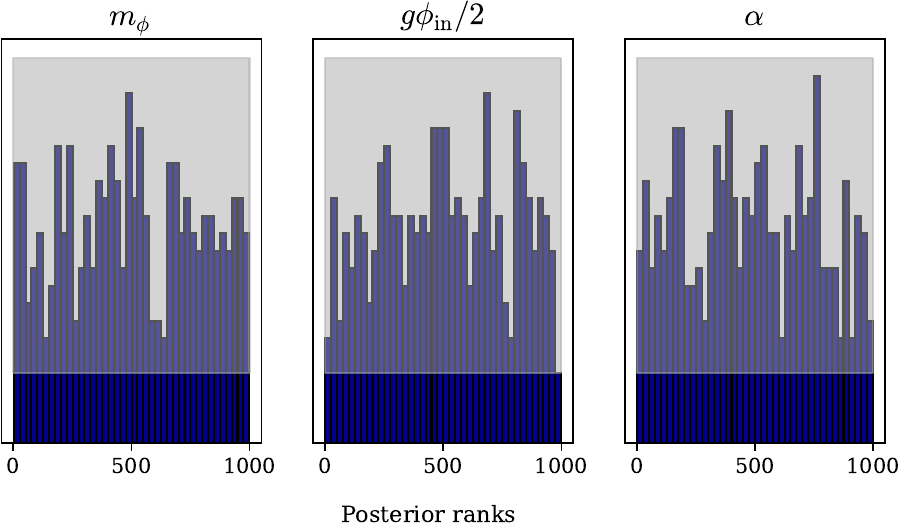} 
    \caption{Distribution of the posterior rank statistics for $m_{\phi}$, $g\phi_{\mathrm{in}}/2$, $\alpha$ for the amortized inference NPE. The shaded area represents the $99\,\%$ confidence level of a uniform distribution. }\label{rank_stats}
\end{figure}

We perform SBC by generating 500 parameter samples from the prior $\boldsymbol{\theta} \sim p(\boldsymbol{\theta})$, where $\boldsymbol{\theta} = (m_{\phi}, g\phi_{\mathrm{in}}/2,\alpha)$ and produce the corresponding $\widehat{C}_{\ell}^{EB}$ simulations. We evaluate the posterior for each $\widehat{C}_{\ell}^{EB}$, draw a 1000-element posterior sample and compute the rank statistics in each dimension (for $m_{\phi}$, $g\phi_{\mathrm{in}}/2$, and $\alpha$). The distribution of the rank statistics, shown in \cref{rank_stats}, confirms a well-calibrated posterior as we do not observe significant deviation from uniformity, with the shaded areas representing the $99\,\%$ confidence level of a uniform distribution.

\begin{figure}
    \centering
    \includegraphics[width=\textwidth]{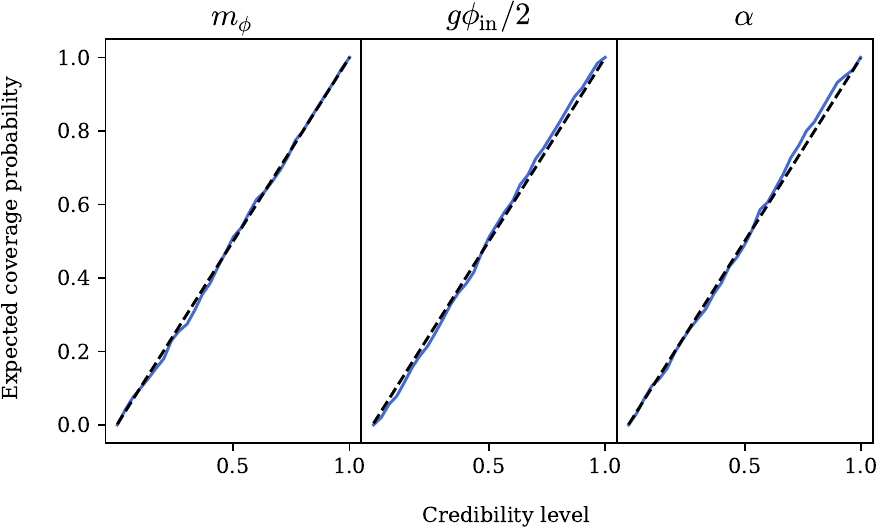} 
    \caption{TARP coverage test: expected coverage probability (ECP) as a function of the credibility level for $m_{\phi}$, $g\phi_{\mathrm{in}}/2$, and $\alpha$. The black dashed line corresponds to the maximum posterior accuracy for which the ECP perfectly matches the credibility level. The blue line represents the outcome of the TARP test for the posterior obtained in \cref{am_inf}.}\label{Tarp}
\end{figure}

As a complement to SBC,  posterior coverage tests represent a necessary and sufficient condition for an accurate posterior. For instance, the \textit{Test of accuracy with random points} (TARP)~\cite{TARP} consists of sampling $\boldsymbol{\theta}_i$ from the prior and producing associated data $\boldsymbol{x}_{i}$. For each simulation $\boldsymbol{x}_{i}$, a set of posterior samples $\boldsymbol{\theta}_s^{(j)}$ is generated from the inferred joint posterior distribution. Assuming a reference point $\theta_r$, we count how many posterior samples lie within a specified radius around $\theta_r$, defining a credible region at a given credibility level. Repeating this over many simulations allows us to compute the expected coverage probability (ECP), i.e., the fraction of simulations in which the true parameter lies within the credible region. For a well-calibrated posterior, the ECP is expected to match the nominal credibility level. \Cref{Tarp} shows the ECP as a function of the credibility level for our three parameters, $m_{\phi}$, $g \phi_{\mathrm{in}}/2$, and $\alpha$, demonstrating that the posterior is well-calibrated, as the credibility levels and ECP are fully consistent.

\bibliographystyle{JHEP}
\bibliography{bibliographie}
\end{document}